\documentclass[12pt]{article}

\usepackage[a4paper, margin=1in]{geometry} 
\usepackage{lmodern} 
\usepackage[T1]{fontenc}
\usepackage{amsmath, amssymb, amsthm} 
\usepackage{graphicx} 
\usepackage{xcolor} 
\usepackage{booktabs} 
\usepackage{titlesec} 
\usepackage{fancyhdr} 
\usepackage{tocloft} 
\usepackage{natbib} 
\usepackage{hyperref,url} 
\usepackage{geometry}
\usepackage{xcolor,inconsolata}
\usepackage{listings}

\definecolor{richblue}{rgb}{0.1, 0.2, 0.7}
\definecolor{darkgray}{rgb}{0.2, 0.4, 0.8}

\definecolor{codebg}{RGB}{245, 245, 245} 
\definecolor{keywordcolor}{RGB}{0, 0, 255} 
\definecolor{stringcolor}{RGB}{163, 21, 21} 
\definecolor{commentcolor}{RGB}{0, 128, 0} 

\hypersetup{
	colorlinks=true,
	linkcolor=richblue,
	urlcolor=richblue,
	citecolor=richblue
}

\titleformat{\section}
{\large\bfseries\color{richblue}}
{\thesection}{1em}{}

\titleformat{\subsection}
{\normalsize\bfseries\color{darkgray}}
{\thesubsection}{1em}{}

\newcommand{\mb}{\mathbf}

\newcommand{\be}{\begin {equation}}
\newcommand{\ee}{\end {equation}}
\newcommand{\beqa}{\begin {eqnarray}}
\newcommand{\eeqa}{\end {eqnarray}}
\newcommand{\DDT}[1]{\frac{d#1}{dt} }
\newcommand{\DDTa}[1]{\frac{d#1}{d\eta} }
\newcommand{\pddt}[1]{ \partial#1/\partial t}
\newcommand{\pddz}[1]{ \partial#1/\partial z}
\newcommand{\Pdd}[2]{\frac{\partial#1}{\partial #2} }
\title{\textbf{\Large\color{richblue} Relativistic electron dynamics in ultra-intense lasers
}}
\author{\large Amol R Holkundkar \\
	\textit{\small{\color{darkgray} Birla Institute of Technology and Science - Pilani, Pilani.}} \\ 
	\texttt{\small{amol.holkundkar@pilani.bits-pilani.ac.in}}}
\date{9th Dec 2024\\ \textbf{\small{Winter School on Intense Laser Science (WiSILS-2024) @ IIT - Jodhpur}}
	}

\begin{document}
	
	\maketitle
	\thispagestyle{empty}
	
	\begin{abstract}
	 These couple of lectures at WiSILS-2024 will focus on the Relativistic charge particle dynamics in ultra-intense laser pulses. We will be learning about the \textit{relativistic equations of motion},\textit{ Radiation-Reaction}, \textit{Thomson scattering},\textit{ Ponderomotive scattering}, and its usage as the potential diagnostic tool. A simulation code \textit{LEADS} will address the relativistic electron dynamics in the intense-laser pulse. 
	\end{abstract}
	
	\tableofcontents
	\newpage

	\section{Relativistic equation of Motion}
 	  Let us first study the basics of the relativistic dynamics of a charge particle interacting with the ultra-intense laser pulse. 
 	  
	\subsection{Lagrangian Formulation}
    The Lagrangian for a relativistic charged particle moving in an electromagnetic field is given by:
    \be
    	\mathcal{L} = -m c^2 \sqrt{1 - \frac{\mathbf{v}^2}{c^2}} + q \mathbf{v} \cdot \mathbf{A} - q \phi,
    \ee
    where:
    \begin{itemize}
    	\item[-] $m$: Mass of the particle,
    	\item[-] $c$: Speed of light in vacuum,
    	\item[-] $\mathbf{v}$: Velocity of the particle,
    	\item[-] $q$: Electric charge of the particle,
    	\item[-] $\phi$: Scalar potential (electric potential),
    	\item[-] $\mathbf{A}$: Vector potential (associated with the magnetic field).
    \end{itemize}

     The term $-m c^2 \sqrt{1 - \frac{\mathbf{v}^2}{c^2}}$ represents the relativistic kinetic term, derived from the proper time interval and the relativistic energy of the particle.
     The term $q \mathbf{v} \cdot \mathbf{A}$ represents the interaction with the magnetic field (via the vector potential $\mathbf{A}$).
     The term $-q \phi$ represents the interaction with the electric field (via the scalar potential $\phi$).
 
	The Euler-Lagrange equations derived from this Lagrangian yield the Lorentz force equation :
	\be
		\frac{d\mathbf{p}}{dt} = q (\mathbf{E} + \mathbf{v} \times \mathbf{B}),
	\ee
	where:
	\begin{itemize}
		\item $\mathbf{p} = \gamma m \mathbf{v}$ is the relativistic momentum,
		\item $\mathbf{E} = -\nabla \phi - \frac{\partial \mathbf{A}}{\partial t}$ is the electric field,
		\item $\mathbf{B} = \nabla \times \mathbf{A}$ is the magnetic field.
	\end{itemize}
	
	This formulation is especially useful in relativistic mechanics and field theory, as it respects the principles of Lorentz invariance and provides a compact way to describe the motion of a charged particle in electromagnetic fields. 
	
  \subsection{Relativistic Lorentz force equation in covariant form}
  The relativistic Lorentz force equation  in 4-vector notation (SI units) is:
  
  \be
  \frac{d p^\mu}{d\tau} = q F^{\mu\nu} u_\nu
  \ee
  
  where:
  
  \begin{itemize}
  	\item \( p^\mu \) is the 4-momentum of the particle:\( p^\mu = \left( \mathcal{E}/c, \mathbf{p} \right) \) 
  	\item \( \mathcal{E} = \gamma m c^2 \) being the total relativistic energy, 
  	\item \( \mathbf{p} \) the 3-momentum, and \( c \) the speed of light.
  	\item \( \tau \) is the proper time of the particle.
  	\item \( q \) is the charge of the particle.
  	\item \( F^{\mu\nu} \) is the electromagnetic field tensor, encapsulating the electric field \( \mathbf{E} \) and the magnetic field \( \mathbf{B} \):
  	\be
  	F^{\mu\nu} = \begin{pmatrix}
  		0 & -E_x/c & -E_y/c & -E_z/c \\
  		E_x/c & 0 & -B_z & B_y \\
  		E_y/c & B_z & 0 & -B_x \\
  		E_z/c & -B_y & B_x & 0
  	\end{pmatrix}.
  	\ee
  	\item \( u_\nu \) is the 4-velocity of the particle: \( u_\nu = \gamma (c, -\mathbf{v}) \), where \( \gamma = 1/\sqrt{1 - v^2/c^2} \).
  \end{itemize}
  
  This equation  describes the relativistic motion of a charged particle under the influence of electromagnetic fields. The term \( q F^{\mu\nu} u_\nu \) represents the force on the particle in 4-dimensional spacetime.

\subsection{0th Component of the Lorentz Force}

We are interested in the 0th component of the Lorentz force law, so we set \( \mu = 0 \):

\be
\frac{d p^0}{d\tau} = q F^{0\nu} u_\nu
\ee

Substitute the components of \( F^{0\nu} \) and \( u_\nu \) into the above equation .

From the field tensor \( F^{\mu\nu} \), we know that:

\be
F^{00} = 0, \quad F^{01} = -\frac{E_x}{c}, \quad F^{02} = -\frac{E_y}{c}, \quad F^{03} = -\frac{E_z}{c}
\ee

Now, for the covariant 4-velocity:

\be
u_\nu = \gamma(c, -\mathbf{v})
\ee

Thus, the components are:

\be
u_0 = \gamma c, \quad u_1 = -\gamma v_x, \quad u_2 = -\gamma v_y, \quad u_3 = -\gamma v_z
\ee

Now, compute \( F^{0\nu} u_\nu \):

\be
F^{0\nu} u_\nu = F^{00} u_0 + F^{01} u_1 + F^{02} u_2 + F^{03} u_3
\ee

Substituting the components:

\be
F^{00} u_0 = 0 \cdot \gamma c = 0
\ee
\be
F^{01} u_1 = -\frac{E_x}{c} \cdot (-\gamma v_x) = \gamma \frac{E_x v_x}{c}
\ee
\be
F^{02} u_2 = -\frac{E_y}{c} \cdot (-\gamma v_y) = \gamma \frac{E_y v_y}{c}
\ee
\be
F^{03} u_3 = -\frac{E_z}{c} \cdot (-\gamma v_z) = \gamma \frac{E_z v_z}{c}
\ee

Thus, we have:

\be
F^{0\nu} u_\nu = \gamma \left( \frac{E_x v_x}{c} + \frac{E_y v_y}{c} + \frac{E_z v_z}{c} \right)
\ee
 
Substituting this expression into the Lorentz force equation :

\be
\frac{d p^0}{d\tau} = q \gamma \left( \frac{E_x v_x}{c} + \frac{E_y v_y}{c} + \frac{E_z v_z}{c} \right)
\ee

This simplifies with $p^0 = \mathcal{E}/c$:

\be
\frac{1}{c}\frac{d\mathcal{E}}{d\tau} = q \gamma \left( \mathbf{E} \cdot \frac{\mathbf{v}}{c} \right)
\ee

\be
 \frac{d\mathcal{E}}{d\tau} = q \gamma \left( \mathbf{E} \cdot \mathbf{v} \right) \implies 
 \gamma \frac{d\mathcal{E}}{dt} = q \gamma \left( \mathbf{E} \cdot \mathbf{v} \right) \implies
  \frac{d\mathcal{E}}{dt} = q \left( \mathbf{E} \cdot \mathbf{v} \right)
\ee

\subsection{Interpretation}

\begin{itemize}
	\item The left-hand side, \( d\mathcal{E}/dt \), represents the rate of change of the particle's energy.
	\item The right-hand side represents the power delivered to the particle by the electric field. The term \( \mathbf{E} \cdot \mathbf{v} \) is the instantaneous power from the electric field.
\end{itemize}

\section{Electromagnetic Fields of a Moving Charge}

The electromagnetic fields of a moving charge, particularly an accelerating charge, are fully described using the \textbf{Lienard-Wiechert potentials}. These fields depend on the position, velocity, and acceleration of the charge and provide a complete solution to Maxwell's equations for a moving point charge.

\subsection{Lienard-Wiechert Potentials}
The electromagnetic fields of a moving point charge are derived from the following potentials:

\begin{itemize}
    \item \textbf{Scalar Potential (\( \phi \))}:
    \be
    \phi(\mathbf{r}, t) = \frac{q}{4 \pi \epsilon_0} \frac{1}{|\mathbf{r} - \mathbf{r}'(t_r)| (1 - \mathbf{n} \cdot \boldsymbol{\beta})}
    \ee

    \item \textbf{Vector Potential (\( \mathbf{A} \))}:
    \be
    \mathbf{A}(\mathbf{r}, t) = \frac{q \boldsymbol{\beta}}{4 \pi \epsilon_0 c} \frac{1}{|\mathbf{r} - \mathbf{r}'(t_r)| (1 - \mathbf{n} \cdot \boldsymbol{\beta})}
    \ee
\end{itemize}

Here:
\begin{itemize}
    \item \( \mathbf{r}'(t_r) \) is the position of the charge at the retarded time \( t_r \) (the time when the signal was emitted from the charge).
    \item \( \boldsymbol{\beta} = \frac{\mathbf{v}}{c} \) is the charge's velocity in units of \( c \).
    \item \( \mathbf{n} = \frac{\mathbf{r} - \mathbf{r}'}{|\mathbf{r} - \mathbf{r}'|} \) is the unit vector from the charge to the observation point.
    \item \( 1 - \mathbf{n} \cdot \boldsymbol{\beta} \) is the relativistic correction factor due to the motion of the charge.
\end{itemize}

\subsection{Fields from Lienard-Wiechert Potentials}
Using these potentials, the electric and magnetic fields of the moving charge are:
 
The total electric field is a combination of two terms: the \textit{Coulomb-like field} and the \textit{radiation field}.

\be
\mathbf{E} = \frac{q}{4 \pi \epsilon_0} \left[ \frac{(1 - \beta^2)\mathbf{n}}{(1 - \mathbf{n} \cdot \boldsymbol{\beta})^3 R^2} + \frac{\mathbf{n} \times \{ (\mathbf{n} - \boldsymbol{\beta}) \times \dot{\boldsymbol{\beta}} \}}{c (1 - \mathbf{n} \cdot \boldsymbol{\beta})^3 R} \right]
\ee

Here:
\begin{itemize}
    \item The \textbf{first term} represents the Coulomb-like field, which depends on the velocity (\( \boldsymbol{\beta} \)) of the charge.
    \item The \textbf{second term} represents the radiation field, which arises from the charge's acceleration (\( \dot{\boldsymbol{\beta}} \)).
\end{itemize}

The magnetic field is related to the electric field as:
\be
\mathbf{B} = \frac{\mathbf{n} \times \mathbf{E}}{c}
\ee

\subsection{What is \( R \)?}

In the context of the electromagnetic fields of a moving charge, \( R \) represents the \textbf{retarded distance} between the charge and the observation point. It is defined as:
\be
R = |\mathbf{r} - \mathbf{r}'(t_r)|
\ee

\subsection*{Explanation}
\begin{enumerate}
    \item \textbf{Observation Point (\( \mathbf{r} \)):}
    The position in space where the electromagnetic field is being measured or observed.
    
    \item \textbf{Position of the Charge (\( \mathbf{r}'(t_r) \)):}
    The location of the charge at the \textit{retarded time} \( t_r \), which accounts for the time it takes for the electromagnetic signal to propagate to the observation point at the speed of light.
    
    \item \textbf{Retarded Distance:}
    The distance between the charge at its position at the retarded time (\( \mathbf{r}'(t_r) \)) and the observation point (\( \mathbf{r} \)).
\end{enumerate}

Mathematically:
\be
R = |\mathbf{r} - \mathbf{r}'(t_r)| = \sqrt{(x - x'(t_r))^2 + (y - y'(t_r))^2 + (z - z'(t_r))^2}.
\ee

\subsection*{Retarded Time (\( t_r \))}
The retarded time \( t_r \) is defined implicitly by the relation:
\be
t_r = t - \frac{R}{c},
\ee
where:
\begin{itemize}
    \item \( t \) is the time of observation,
    \item \( c \) is the speed of light.
\end{itemize}

This ensures causality, as the fields at the observation point depend on the charge's state at an earlier time, accounting for the finite speed of electromagnetic signal propagation.

\subsection{Physical Insights}
\begin{enumerate}
    \item \textbf{Near Field (Coulomb-like term):}
    \begin{itemize}
        \item Dominates at distances close to the charge.
        \item Falls off as \( 1/R^2 \).
        \item Depends on the instantaneous velocity of the charge.
    \end{itemize}

    \item \textbf{Far Field (Radiation term):}
    \begin{itemize}
        \item Dominates at large distances.
        \item Falls off as \( 1/R \), indicating radiative energy transfer.
        \item Depends on the instantaneous acceleration of the charge.
    \end{itemize}

    \item \textbf{Angular Dependence:}
    \begin{itemize}
        \item The radiation field is strongest in directions perpendicular to the acceleration vector.
        \item The fields are zero along the direction of motion for linear acceleration.
    \end{itemize}
\end{enumerate}

\subsection{Special Cases}
\begin{enumerate}
    \item \textbf{Uniform Motion (\( \mathbf{a} = 0 \)):}
    \begin{itemize}
        \item If the charge moves at a constant velocity, there is no radiation field.
        \item The fields simplify to a Coulomb-like field modified by relativistic effects (contracted in the direction of motion).
    \end{itemize}

    \item \textbf{Accelerated Motion (\( \mathbf{a} \neq 0 \)):}
    \begin{itemize}
        \item Radiation fields appear.
        \item The fields are time-dependent and carry energy away as electromagnetic waves.
    \end{itemize}

    \item \textbf{Oscillatory Motion:}
    \begin{itemize}
        \item Oscillating charges produce periodic radiation, such as dipole or quadrupole radiation patterns, depending on the symmetry of the motion.
    \end{itemize}
\end{enumerate}
	
\section{Power Radiated Per Unit Solid Angle (Relativistic Case)}

\subsection{Relativistic Electromagnetic Field}
The far-field electric field \( \mathbf{E}_{\text{rad}} \) for a relativistically moving charge \( q \) is given by:
\be
\mathbf{E}_{\text{rad}} = \frac{q}{4 \pi \epsilon_0 c} \frac{\mathbf{n} \times \left[(\mathbf{n} - \boldsymbol{\beta}) \times \dot{\boldsymbol{\beta}}\right]}{\left(1 - \mathbf{n} \cdot \boldsymbol{\beta}\right)^3 R},
\ee
where:
\begin{itemize}
    \item \( \mathbf{n} \) is the unit vector pointing from the charge to the observation point,
    \item \( \boldsymbol{\beta} = \frac{\mathbf{v}}{c} \) is the velocity of the charge (in units of \( c \)),
    \item \( \dot{\boldsymbol{\beta}} = \frac{\mathbf{a}}{c} \) is the acceleration of the charge (in units of \( c \)),
    \item \( 1 - \mathbf{n} \cdot \boldsymbol{\beta} \) is the Doppler factor, which accounts for relativistic corrections,
    \item \( R \) is the distance to the observation point.
\end{itemize}

The corresponding magnetic field is:
\be
\mathbf{B}_{\text{rad}} = \frac{\mathbf{n} \times \mathbf{E}_{\text{rad}}}{c}.
\ee

\subsection{Poynting Vector}
The Poynting vector, representing the radiated energy flux per unit area, is:
\be
\mathbf{S} = \frac{1}{\mu_0} (\mathbf{E}_{\text{rad}} \times \mathbf{B}_{\text{rad}}).
\ee
Substituting \( \mathbf{B}_{\text{rad}} \) into \( \mathbf{S} \), we get:
\be
\mathbf{S} = \frac{1}{\mu_0 c} |\mathbf{E}_{\text{rad}}|^2 \mathbf{n}.
\ee

The energy per unit area per unit time detected at the obsrevation point at time $t$ when 
the radiation is emitted at $t_r = t - R/c$.
\be
 [\mathbf{S\cdot \mathbf{n}]_\text{ret}} = \frac{1}{\mu_0 c} |\mathbf{E}_{\text{rad}}|^2.
\ee

Substituting the relativistic \( \mathbf{E}_{\text{rad}} \), we obtain:
\be
|\mathbf{E}_{\text{rad}}|^2 = \left( \frac{q}{4 \pi \epsilon_0 c} \right)^2 \frac{|\mathbf{n} \times \left[(\mathbf{n} - \boldsymbol{\beta}) \times \dot{\boldsymbol{\beta}}\right]|^2}{\left(1 - \mathbf{n} \cdot \boldsymbol{\beta}\right)^6 R^2}.
\ee

The energy radiated per unit area from $t_{r1}$ to $t_{r2}$ is:
\be \mathcal{E}_{radiated} = \int_{t_{r1}}^{t_{r2}} [\mathbf{S\cdot \mathbf{n}]_\text{ret}} dt  =  \int_{t_{r1}}^{t_{r2}} [\mathbf{S\cdot \mathbf{n}]_\text{ret}} \frac{dt}{dt_r} dt_r\ee

as $t_r = t - R/c \implies dt/dt_r = 1 - \mathbf{n}\cdot\boldsymbol{\beta}$ above equation  reduces to (please note that $\mathbf{R}$ is the vector from the source to the observation point, however when we are doing the measurement all the quantites are calculated in Lab frame and hence a negative sign will come in the expression):
\be \mathcal{E}_{radiated} =  \int_{t_{r1}}^{t_{r2}} \Big\{  [\mathbf{S\cdot \mathbf{n}]_\text{ret}} [1 - \mathbf{n}\cdot\boldsymbol{\beta}] \Big\} dt_r\ee

Now the term in the $\{\cdots\}$ is actually the power radiated per unit area. 

\subsection{Power Radiated Per Unit Solid Angle}
From the above analysis the power radiated per unit solid angle (at retarded time) is related to the Poynting vector by:
\be
\frac{dP(t_r)}{d\Omega} = R^2 |\mathbf{S}\cdot \mathbf{n}|_\text{ret} [1 - \mathbf{n}\cdot\boldsymbol{\beta}] \quad\quad [\because dA = R^2 d\Omega = R^2 \sin\theta d\theta d\phi] .
\ee

Using the expression fo
r \( |\mathbf{E}_{\text{rad}}|^2 \), we get:
\be
\frac{dP}{d\Omega} = \frac{1}{\mu_0 c} \left( \frac{q}{4 \pi \epsilon_0 c} \right)^2 \frac{|\mathbf{n} \times \left[(\mathbf{n} - \boldsymbol{\beta}) \times \dot{\boldsymbol{\beta}}\right]|^2}{\left(1 - \mathbf{n} \cdot \boldsymbol{\beta}\right)^5}.
\ee

Simplifying further using \( \mu_0 c^2 = \frac{1}{\epsilon_0} \):
\be
\frac{dP(t_r)}{d\Omega} = \frac{q^2}{16 \pi^2 \epsilon_0 c} \frac{|\mathbf{n} \times [(\mathbf{n} - \boldsymbol{\beta}) \times \dot{\boldsymbol{\beta}}]|^2}{\left(1 - \mathbf{n} \cdot \boldsymbol{\beta}\right)^5}.
\ee

However, for the power radiated per unit solid angle as evaluated at the time $t$ is then given by:

\be
\frac{dP(t)}{d\Omega} = \frac{q^2}{16 \pi^2 \epsilon_0 c} \left|\frac{\mathbf{n} \times [(\mathbf{n} - \boldsymbol{\beta}) \times \dot{\boldsymbol{\beta}}]}{\left(1 - \mathbf{n} \cdot \boldsymbol{\beta}\right)^3}\right|^2
\equiv |V(t)|^2 \quad \text{say}\ee

\subsection{Angular Dependence}
The term \( |\mathbf{n} \times [(\mathbf{n} - \boldsymbol{\beta}) \times \dot{\boldsymbol{\beta}}]|^2 \) describes the angular dependence of the radiation:
\begin{itemize}
    \item The factor \( \left(1 - \mathbf{n} \cdot \boldsymbol{\beta}\right)^{-5} \) leads to strong collimation of radiation along the direction of motion (relativistic beaming).
    \item The radiation is strongest perpendicular to the direction of acceleration, as in the non-relativistic case.
\end{itemize}

\begin{figure}[!t]
   \centering \includegraphics[width=0.7\textwidth]{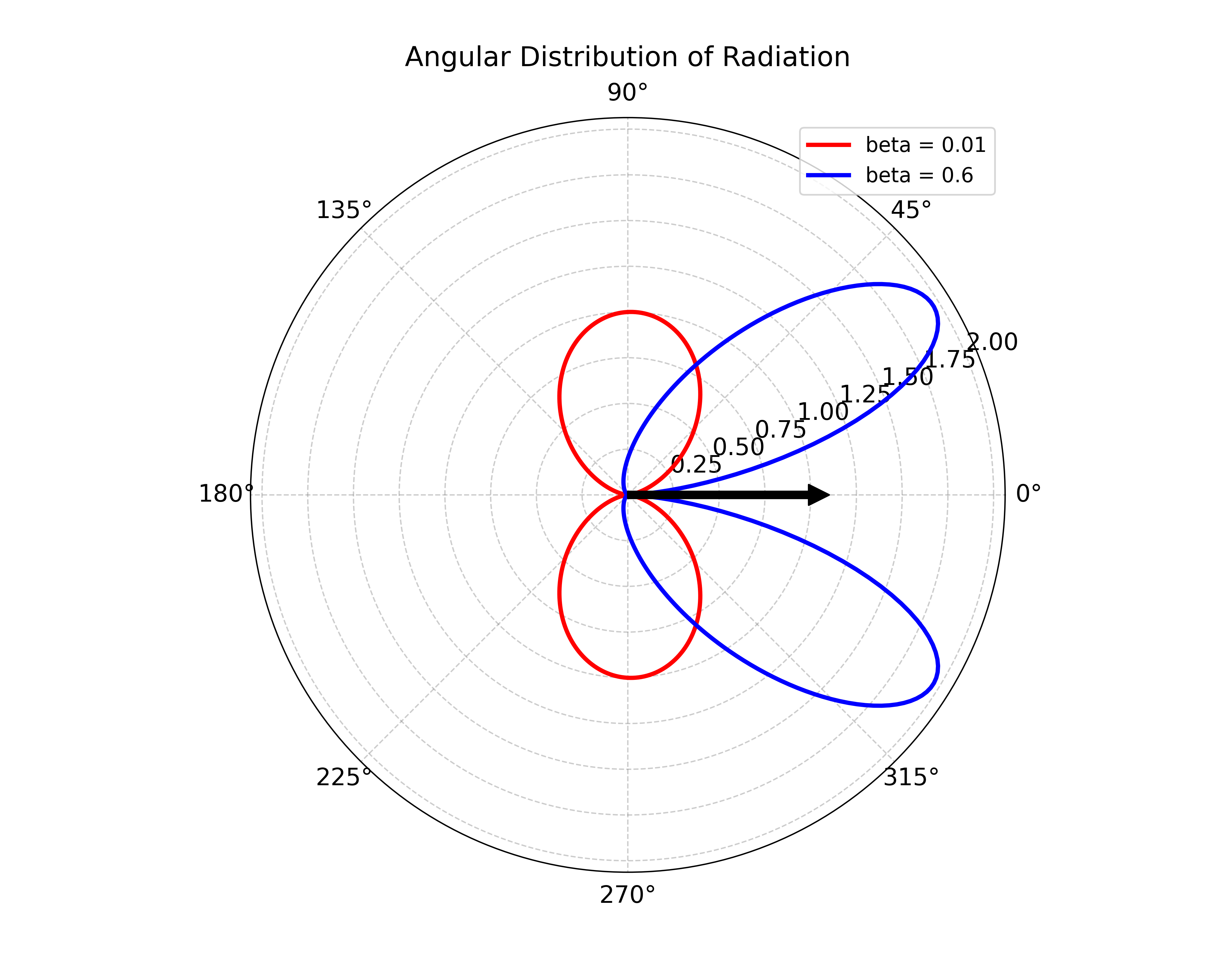}
    \caption{Angular distribution of the power radiated for $\beta \ll 1$ and $\beta \rightarrow 1$.}
\end{figure}

In the special case $\boldsymbol{\beta} || \boldsymbol{\dot{\beta}}$ and $\mathbf{n}\cdot\boldsymbol{\beta} = \beta \cos\theta$, the above equation  simplifies to:
\be
\boxed{\frac{dP(t_r)}{d\Omega} = \frac{q^2 \dot{\beta}^2 \sin^2 \theta}{16 \pi^2 \epsilon_0 c (1 - \beta \cos \theta)^5}}
\ee

From this expression it is clear that for non-relativistic cases $\beta \ll 1$ the maximum power will be emitted along $\theta = \pi/2$, however for relativistic cases $\beta \sim 1$ the radiation will be emitted along the direction of the electron motion. 

\subsection{Total power radiated : Larmor Formula}

Total power radiated can be calcuated by integrating the $dP/d\Omega$ as:

\be 
 P(t_r) = \frac{q^2 \dot{\beta}^2}{16 \pi^2 \epsilon_0 c} \int_0^{2\pi} d\phi \int_0^\pi \frac{\sin^2\theta}{(1 - \beta \cos \theta)^5}
 \sin\theta d\theta = \frac{q^2 \dot{\beta}^2}{16 \pi^2 \epsilon_0 c}\times 2 \pi \times \frac{4}{3} \frac{1}{(1-\beta^2)^3} \ee

As we know $\gamma = 1/\sqrt{1-\beta^2}$	and hence above equations simplifies to give us the total power radiated, which is refererred as the Larmor Formula. 

\be P(t_r) = \frac{q^2 \dot{\beta}^2 \gamma^6}{6 \pi \epsilon_0 c} =  \frac{\mu_0 q^2}{6 \pi c} \dot{v}^2 \gamma^6 \ee

\section{Frequency distribution of the emitted radiation}
 
 Earlier we obtained the power radiated per unit solid angle as:
 
\be
\frac{dP(t)}{d\Omega} = \frac{q^2}{16 \pi^2 \epsilon_0 c} \left|\frac{\mathbf{n} \times [(\mathbf{n} - \boldsymbol{\beta}) \times \dot{\boldsymbol{\beta}}]}{\left(1 - \mathbf{n} \cdot \boldsymbol{\beta}\right)^3}\right|^2
\equiv |V(t)|^2 \quad \text{say}\ee
  
Hence the energy radiated per unit solid angle is then given by using Parseval's theorem:
\[
\frac{dW}{d\Omega} = \int |V(t)|^2 \, dt = \int |\tilde{V}(\omega)|^2 \, d\omega,
\]
where \( \tilde{V}(\omega) \) is the Fourier transform of \( V(t) \), defined as:
\[
\tilde{V}(\omega) = \frac{1}{\sqrt{2\pi}}\int V(t) e^{i\omega t} \, dt.
\]

\[
V(t) = \frac{1}{\sqrt{2\pi}}\int \tilde{V}(\omega) e^{-i\omega t} \, d\omega.
\]
 
The energy radiated per unit frequency per unit solid angle is then given by:
\[
\frac{d^2I}{d\omega d\Omega} = |\tilde{V}(\omega)|^2.
\]

 The energy radiated per unit frequency and unit solid angle becomes after using previous expressions:
 
  \be
 \frac{d^2I}{d\omega d\Omega} = \frac{q^2}{16\pi^3 \epsilon_0 c} \left| \int_{-\infty}^\infty \frac{\hat{n} \times \left[ (\hat{n} - \vec{\beta}) \times \dot{\vec{\beta}} \right]}{(1 - \hat{n} \cdot \vec{\beta})^3} e^{i\omega t} \, dt \right|^2.
 \ee
 
 Substituting, $t = t_r + R/c $ and $\int dt = \int dt_r (dt/dt_r) = \int (1 - \hat{n} \cdot \vec{\beta}) dt_r$ the above expression simplifies to,
 
 \be
 \frac{d^2I}{d\omega d\Omega} = \frac{q^2}{16\pi^3 \epsilon_0 c} \left| \int_{-\infty}^\infty \frac{\hat{n} \times \left[ (\hat{n} - \vec{\beta}) \times \dot{\vec{\beta}} \right]}{(1 - \hat{n} \cdot \vec{\beta})^2} e^{i\omega(t_r + R/c)} \, dt_r \right|^2.
 \ee
 
 \subsection{Explanation of Terms}
 
 \begin{itemize}
 	\item The cross-product \( \hat{n} \times \left[ (\hat{n} - \vec{\beta}) \times \dot{\vec{\beta}} \right] \) describes the angular and polarization dependence of the radiation.
 	\item The term \( (1 - \hat{n} \cdot \vec{\beta})^2 \) accounts for relativistic time dilation and Doppler shifts.
 	\item The exponential factor \( e^{i\omega(t_r + R/c)} \) introduces the frequency dependence, linking the particle's motion to the observed radiation.
 	\item The integration over \( t_r \) connects the particle's trajectory to the radiation field.
 \end{itemize}

\section{Introduction to the Radiation Reaction}

Radiation reaction is a phenomenon that occurs when an accelerating charged particle emits electromagnetic radiation. As the particle radiates energy, it experiences a recoil force that alters its motion. This concept lies at the intersection of classical electrodynamics and modern physics and is essential for understanding processes involving high-energy particles, astrophysics, and accelerator physics.

\subsection{When RR is important}

As we saw previously, the total power radiated by the charge particle is given by:

\be P(t_r) =    \frac{\mu_0 q^2}{6 \pi c} a^2 \gamma^6 \ee
where, $a$ is the acceleration of the particle. The energy radiated by the charge particle in time $T$ is then 
given by:
\be \mathcal{E}_{rad} =  \frac{\mu_0 q^2}{6 \pi c} a^2 \gamma^6 \quad T \ee

If, $\mathcal{E}_{rad} \ll \mathcal{E}_0$ where $\mathcal{E}_0$ is the relevant energy scale of the particle then
in that case the RR is not relevant and one can ignore the same for all practical purposes. The specification of the relevant energy scale $\mathcal{E}_0$ needs a small care. If we assume the particle is going through the continuous acceleration, in that case the energy of the particle after time $T$ would be:
\be \mathcal{E}_0 \sim m_0 (aT)^2  \ee
Therefore, the RR is not important if :
\be \frac{\mu_0 q^2}{6 \pi c} a^2 \gamma^6 \quad T \ll  m_0 (aT)^2 \implies T \gg \frac{\mu_0 q^2}{6 \pi c m_0}  \gamma^6 \equiv \tau  \ee
The characteristic time-scale where RR is important for \textbf{electron dynamics} is then given by:
\be \tau = \frac{\mu_0}{6\pi c} \frac{q^2}{m_0}  \gamma^6 \sim 6.27\times 10^{-24} \gamma^6\ \ \text{sec}\ee
 
\subsection{Radiation Reaction Force from Conservation of Energy}

The radiation reaction force can be derived from the principle of conservation of energy, considering the power radiated as electromagnetic waves by an accelerating charge (Larmor power).

\subsection*{Radiated Power (Larmor Formula)}

The power radiated by a non-relativistic charge is given by the Larmor formula:
\be
P = \frac{\mu_0 q^2 a^2}{6\pi c}
\ee
where:
\begin{itemize}
    \item \(a\) is the acceleration of the charge,
    \item \(q\) is the charge,
    \item \(\mu_0\) is the permeability of free space,
    \item \(c\) is the speed of light.
\end{itemize}

\subsection{Work-Energy Principle and Abraham-Lorentz equation }
The equation  of motion with the RR is given by:
$$ m\mathbf{a} = \mathbf{F}_{\text{ext}}  + \mathbf{F}_{\text{rad}} $$
In order to determine the form of the $\mathbf{F}_{\text{rad}}$, we can say that the work done by this
force is equal to the negative of the energy radiated away:

The work done by a force is related to the rate of change of the particle's kinetic energy:
\be
\text{Work done by forces} = \text{Change in kinetic energy}.
\ee
 
\be \int_{t_1}^{t_2}  \mathbf{F}_{\text{rad}} \cdot \mathbf{v}\ \ dt = - \int_{t_1}^{t_2} \frac{\mu_0 q^2}{6\pi c} (\dot{\mathbf{v}} \cdot \dot{\mathbf{v}})\ \ dt = -\frac{\mu_0 q^2}{6\pi c} \Big\{ \big[\dot{\mathbf{v}} \cdot \mathbf{v}\big]_{t_1}^{t_2} - \int_{t_1}^{t_2} \ddot{\mathbf{v}} \cdot \mathbf{v}\ dt  \Big\}   \ee

For the circular motion or a periodic motion $\mathbf{\dot{v}} \perp \mathbf{v}$ and hence the first term in the [...] is zero. For periodic motion the value in the bracket would be same at $t_1$ and $t_2$ and hence it will be zero. Now comparing the RHS and LHS, the form of the $\mathbf{F}_{\text{rad}}$ can be deduced as: 

\be
\mathbf{F}_{\text{rad}} = \frac{\mu_0 q^2}{6\pi c} \frac{d\mathbf{a}}{dt}
\ee
where \( d\mathbf{a}/dt \) is the time derivative of acceleration, often called the ``jerk.''

The radiated power represents the energy lost by the charge due to radiation. To conserve energy, this must come from an additional force opposing the motion, called the \textbf{radiation reaction force} (\(\mathbf{F}_{\text{rad}}\)).
 
\begin{itemize}
    \item The force depends on the rate of change of acceleration, which is a third-order time derivative of position.
    \item It is very small in magnitude because the prefactor \( \mu_0 q^2/(6\pi c) \) is extremely small, reflecting the fact that radiation reaction effects are typically negligible unless accelerations are very large (e.g., in particle accelerators or astrophysical processes).
\end{itemize}

Now, the equation  of the motion will be then modified as:
$$ \boxed{ m (\mathbf{\dot{v}} - \tau \mathbf{\ddot{v}}) = \mathbf{F}_{ext} }$$ 
where, $\tau =\mu_0 e^2/(6 \pi c m_0)$ for non-relativisitc case $\gamma \sim 1$. This equation  is referred as the 
\textbf{Abraham-Lorentz equation } of motion which approximates the effect of RR. 

The Abraham–Lorentz equation  describes the radiation reaction force on a charged particle due to the electromagnetic radiation it emits when it is accelerated. This reaction arises because the particle loses energy to radiation, which must impact its motion.

\subsection{Key Features of the Abraham–Lorentz equation }
\begin{enumerate}
    \item \textbf{Radiation Reaction Force}: The term \( \frac{\mu_0 q^2}{6 \pi c} \dddot{\mathbf{x}} \) represents the radiation reaction force. It shows that the particle’s own emission of radiation feeds back into its motion.
    \item \textbf{Runaway Solutions}: The equation  predicts unphysical ``runaway" solutions, where the particle's acceleration grows exponentially even in the absence of an external force. This is a fundamental issue of the equation  in its classical form.
    \item \textbf{Preacceleration}: Another issue is the prediction of ``preacceleration," where the particle seems to anticipate an external force and begins moving before the force is applied.
\end{enumerate}

\subsection{What do we mean by Runaway Solutions and Pre-Acceleration?}
As we learned the Abraham-Lorentz equation  is given by:
$$ m (\mathbf{\dot{v}} - \tau \mathbf{\ddot{v}}) = \mathbf{F}_{ext} $$ 
Now, in the absence of any externally applied force $\mathbf{F}_{ext} = 0$ the above equation  is modified as:
$$ \mathbf{\dot{v}} - \tau \mathbf{\ddot{v}} = 0 \implies \frac{d\mathbf{a}}{dt} = \frac{\mathbf{a}}{\tau}\implies \mathbf{a}(t) = \mathbf{a_0} \text{e}^{t/\tau} $$
where, $\mathbf{a}_0$ is the acceleration of the particle at $t = 0$. The above equations shows that the acceleration of the particle 
grows exponentially without application of any applied force, and hence it is referred as the ``Runaway" solutions. 

The Runaway solutions can be remedied by setting $\mathbf{a}_0 = 0$ but it gives rise to the so called ``Pre-Acceleation". 

\subsection*{Example}

With the inclusion of the radiation reaction force, Newton’s second law for a charged particle becomes
\be
\mathbf{a} = \tau \dot{\mathbf{a}} + \frac{\mathbf{F}}{m},
\ee
where \( \mathbf{F} = \mathbf{F}_{ext} \) is the external force acting on the particle.
  
\paragraph{} 

\noindent
\textbf{(a)} A particle is subjected to a constant force \( F \), beginning at time \( t = 0 \) and lasting until time \( T \). Find the most general solution \( a(t) \) to the equation  of motion in each of the three periods: 
\begin{itemize}
	\item[(i)] \( t < 0 \),
	\item[(ii)] \( 0 < t < T \),
	\item[(iii)] \( t > T \).
\end{itemize}

The equation  of motion is:
\be
a - \tau \dot{a} = \frac{F}{m}.
\ee

This is a first-order linear differential equation . The general solution is:
\be
a(t) = A e^{t/\tau} + \frac{F}{m},
\ee
where \(A\) is an integration constant.

We now consider the three regions:

(i) Region \(t < 0\):  
Before \(t = 0\), no force acts on the particle (\(F = 0\)). The equation  reduces to:
\be
a - \tau \dot{a} = 0.
\ee
The solution is:
\be
a(t) = A_1 e^{t/\tau},
\ee
where \(A_1\) is a constant.

(ii) Region \(0 < t < T\):  
A constant force \(F\) acts on the particle. The equation  is:
\be
a - \tau \dot{a} = \frac{F}{m}.
\ee
The solution is:
\be
a(t) = A_2 e^{t/\tau} + \frac{F}{m},
\ee
where \(A_2\) is a constant.

(iii) Region \(t > T\):  
After \(t = T\), the force is removed (\(F = 0\)). The equation  reduces to:
\be
a - \tau \dot{a} = 0.
\ee
The solution is:
\be
a(t) = A_3 e^{t/\tau},
\ee
where \(A_3\) is a constant.

\paragraph{}
\noindent
\textbf{(b)} Impose the continuity condition \( a(t) \) at \( t = 0 \) and \( t = T \). Show that you can \emph{either} eliminate the runaway in region (iii) \emph{or} avoid preacceleration in region (i), but not both.

At \(t = 0\):  
The acceleration \(a(t)\) must be continuous, so:
\be
A_1 e^{0/\tau} = A_2 e^{0/\tau} + \frac{F}{m}.
\ee
Simplifying:
\be
A_1 = A_2 + \frac{F}{m}.
\ee

At \(t = T\):  
The acceleration \(a(t)\) must again be continuous:
\be
A_2 e^{T/\tau} + \frac{F}{m} = A_3 e^{T/\tau}.
\ee
Simplifying:
\be
A_3 = A_2 + \frac{F}{m} e^{-T/\tau}.
\ee

\paragraph{}
\noindent
\textbf{Runaway and Preacceleration:}  
- Runaway: If \(A_3 \neq 0\), the acceleration \(a(t)\) for \(t > T\) diverges exponentially as \(t \to \infty\).
- Preacceleration: If \(A_1 \neq 0\), there is nonzero acceleration before the force is applied (\(t < 0\)).

To avoid these behaviors:  
- Eliminate runaway (\(t > T\)): Set \(A_3 = 0\), which gives:
\be
A_2 = -\frac{F}{m} e^{-T/\tau}.
\ee
- Avoid preacceleration (\(t < 0\)): Set \(A_1 = 0\), which gives:
\be
A_2 = -\frac{F}{m}.
\ee

However, both conditions cannot be satisfied simultaneously, as they impose conflicting requirements on \(A_2\). This highlights the paradox of the radiation reaction force. 

\subsection*{Velocity in the Three Regions}

To calculate the velocity \(v(t)\) in each region, we integrate the acceleration \(a(t)\) over time.

\subsection*{Region (i): \(t < 0\)}

In this region, \(F = 0\), so the acceleration is given by:
\be
a(t) = A_1 e^{t / \tau}.
\ee
The velocity \(v(t)\) is obtained by integrating \(a(t)\):
\be
v(t) = \int a(t) \, dt = \int A_1 e^{t / \tau} \, dt.
\ee
This gives:
\be
v(t) = A_1 \tau e^{t / \tau} + C_1,
\ee
where \(C_1\) is a constant of integration.

\subsection*{Region (ii): \(0 < t < T\)}

In this region, \(F \neq 0\), so the acceleration is:
\be
a(t) = A_2 e^{t / \tau} + \frac{F}{m}.
\ee
Integrating \(a(t)\) to find \(v(t)\):
\be
v(t) = \int \left( A_2 e^{t / \tau} + \frac{F}{m} \right) \, dt.
\ee
This gives:
\be
v(t) = A_2 \tau e^{t / \tau} + \frac{F}{m} t + C_2,
\ee
where \(C_2\) is a constant of integration.

\subsection*{Region (iii): \(t > T\)}

In this region, \(F = 0\), so the acceleration is:
\be
a(t) = A_3 e^{t / \tau}.
\ee
The velocity \(v(t)\) is:
\be
v(t) = \int a(t) \, dt = \int A_3 e^{t / \tau} \, dt.
\ee
This gives:
\be
v(t) = A_3 \tau e^{t / \tau} + C_3,
\ee
where \(C_3\) is a constant of integration.

It should be noted that for the uncharged particle $\tau = 0$ and hence we retrieve the Newton's equation  
of motion. 

\subsection{Lorentz-Abraham-Dirac (LAD) equation }

The Lorentz-Abraham-Dirac (LAD) equation  describes the motion of a charged particle under the influence of external forces and its self-radiation (radiation reaction). The equation  is derived from classical electrodynamics and models the recoil force on a charged particle as it radiates energy due to acceleration. This LAD equation is the relativistic generalization of the Abraham-Lorentz equation. 

\subsection*{LAD Equation : Relativistic Form of the Abraham-Lorentz equation}
Earlier we obtained the non-relativistic version of the RR as is given as:

\be m\mathbf{a} = \mathbf{F}_{\text{ext}}  + \mathbf{F}_{\text{rad}} \ee
where,
\be
\mathbf{F}_{\text{rad}} =  \frac{q^2}{6 \pi \epsilon_0 c^3} \frac{d\mathbf{a}}{dt}
\label{nonrelarad}
\ee

The relativistic equation of motion can then be written as: 
\be m \frac{d u^\mu}{d \tau} = F^\mu_{\text{ext}} + F^\mu_{\text{rad}} \ee
It should be more general and hence in the limit $v \ll c$ we should get results from AL equation. It can be easily proved that in the non-relativistic limit the expression,  
\be F^\mu_{\text{rad}} =  \frac{q^2}{6 \pi \epsilon_0 c^3} g^\mu \quad;\quad g^\mu \equiv \frac{d^2 u^\mu}{d \tau^2} \ee 
is same as equation \ref{nonrelarad}. 

In the relativistic scenario the 4-acceleration (force) should always be orthogonal to the 4-velocity i.e. $a^\mu u_\mu = 0$. This comes from the fact that $u^\mu u_\mu = c^2$ and hence
\be \frac{d(u^\mu u_\mu)}{d\tau} = u_\mu \frac{du^\mu}{d\tau} + u^\mu \frac{du_\mu}{d\tau} = 2 a^\mu u_\mu = 0\ee

However, the $g^\mu$ as defined is not satisfying this condition, and hence we can modify the $g^\mu$ as:
\be g^\mu = \frac{d^2 u^\mu}{d \tau^2} + \delta u^\mu\ee 
and then demand $g^\mu u_\mu = 0$ 
\be g^\mu u_\mu = \frac{d^2 u^\mu}{d \tau^2} u_\mu  + \delta u^\mu u_\mu = 0 \ee
\be \delta  = -\frac{1}{c^2}  \frac{d^2 u^\nu}{d \tau^2} u_\nu \ee
where, $u^\nu u_\nu = c^2$ is used. Let us simplify this further and consider: 
\be \frac{d}{d\tau}\Big( u_\nu \frac{d u^\nu}{d \tau} \Big) = u_\nu \frac{d^2 u^\nu}{d \tau^2} +  \frac{d u_\nu}{d \tau} \frac{d u^\nu}{d \tau} = 0 \ee 
because $a^\nu u_\nu = 0$ is anyway satisfied (LHS of above equation), this implies:
\be \frac{1}{c^2}\frac{d u_\nu}{d \tau} \frac{d u^\nu}{d \tau} = - \frac{1}{c^2} u_\nu \frac{d^2 u^\nu}{d \tau^2} = \delta \ee 

In the relativistic framework, the equation  is then finally written as:
\be
m \frac{d u^\mu}{d \tau} = F^\mu_{\text{ext}} + \frac{q^2}{6 \pi \epsilon_0 c^3} \left( \frac{d^2 u^\mu}{d \tau^2} + \frac{u^\mu}{c^2} \frac{d u_\nu}{d \tau} \frac{d u^\nu}{d \tau} \right),
\ee
where:
\begin{itemize}
	\item \(m\) is the rest mass of the particle,
	\item \(q\) is the charge of the particle,
	\item \(u^\mu = \gamma(c, \vec{v})\) is the 4-velocity (\(\gamma\) is the Lorentz factor),
	\item \(\tau\) is the proper time of the particle,
	\item \(F^\mu_{\text{ext}}\) is the external 4-force acting on the particle.
\end{itemize}

The radiation reaction term consists of:
\begin{enumerate}
	\item  \(\frac{d^2 u^\mu}{d \tau^2}\), which accounts for changes in the particle's acceleration due to radiated energy.
	\item \textbf{Projection term:} \(u^\mu \frac{d u_\nu}{d \tau} \frac{d u^\nu}{d \tau}\), which ensures orthogonality of the 4-velocity and 4-acceleration, maintaining consistency with relativity.
\end{enumerate}

\subsection*{Non-Relativistic Limit}

The non-relativistic limit of the Lorentz-Abraham-Dirac (LAD) equation  is achieved by simplifying the relativistic form under the assumption that the particle's velocity \(v\) is much smaller than the speed of light \(c\) (\(v \ll c\)).
 
The relativistic LAD equation  is:
\be
m \frac{d u^\mu}{d \tau} = F^\mu_{\text{ext}} + \frac{q^2}{6 \pi \epsilon_0 c^3} \left( \frac{d^2 u^\mu}{d \tau^2} + \frac{u^\mu}{c^2} \frac{d u_\nu}{d \tau} \frac{d u^\nu}{d \tau} \right),
\ee
where:
\begin{itemize}
	\item \(u^\mu = \gamma(c, \vec{v})\) is the 4-velocity,
	\item \(\frac{d u^\mu}{d \tau}\) is the 4-acceleration,
	\item \(\tau\) is the proper time.
\end{itemize}

The two main steps to achieve the non-relativistic limit involve simplifying:
\begin{enumerate}
	\item Proper time \(\tau\) in terms of coordinate time \(t\),
	\item Components of the 4-velocity \(u^\mu\) and 4-acceleration \(\frac{d u^\mu}{d \tau}\).
\end{enumerate}

The proper time \(\tau\) relates to the coordinate time \(t\) via the Lorentz factor \(\gamma\):
\be
d\tau = \frac{dt}{\gamma}, \quad \gamma = \frac{1}{\sqrt{1 - v^2/c^2}}.
\ee
In the non-relativistic limit (\(v \ll c\)):
\be
\gamma \approx 1, \quad d\tau \approx dt.
\ee
This allows us to replace derivatives with respect to \(\tau\) with derivatives with respect to \(t\):
\be
\frac{d}{d \tau} \approx \frac{d}{d t}.
\ee

The 4-velocity \(u^\mu = (\gamma c, \gamma \vec{v})\) simplifies in the non-relativistic limit:
\be
u^\mu \approx (c, \vec{v}),
\ee
since \(\gamma \approx 1\).

The 4-acceleration \(\frac{d u^\mu}{d \tau}\) has components:
\be
\frac{d u^\mu}{d \tau} = \left( \frac{d (\gamma c)}{d \tau}, \frac{d (\gamma \vec{v})}{d \tau} \right).
\ee
For \(v \ll c\), we approximate:
\begin{itemize}
	\item The temporal component \(\frac{d (\gamma c)}{d \tau}\) becomes negligible.
	\item The spatial component reduces to the usual 3-acceleration, \(\frac{d \vec{v}}{d t}\).
\end{itemize}

The radiation reaction term in the relativistic LAD equation  is:
\be
F^\mu_{\text{rad}} =\frac{q^2}{6 \pi \epsilon_0 c^3} \left( \frac{d^2 u^\mu}{d \tau^2} + \frac{u^\mu}{c^2} \frac{d u_\nu}{d \tau} \frac{d u^\nu}{d \tau} \right) 
\ee
In the non-relativistic limit:
\begin{enumerate}
	\item The dominant contribution comes from \(\frac{d^2 u^\mu}{d \tau^2}\), as the projection term \(u^\mu \frac{d u_\nu}{d \tau} \frac{d u^\nu}{d \tau}\) involves \(v^2/c^2\), which is small and can be neglected.
	\item The spatial component of \(\frac{d^2 u^\mu}{d \tau^2}\) simplifies to the second derivative of velocity, \(\frac{d^2 \vec{v}}{d t^2}\).
\end{enumerate}

Thus, the radiation reaction force simplifies to:
\be
\vec{F}_{\text{rad}} = \frac{q^2}{6 \pi \epsilon_0 c^3} \frac{d^2 \vec{v}}{d t^2}.
\ee

\subsection*{Final Non-Relativistic LAD equation }

Substituting the simplified expressions into the relativistic LAD equation  yields:
\be
m \frac{d \vec{v}}{d t} = \vec{F}_{\text{ext}} + \frac{q^2}{6 \pi \epsilon_0 c^3} \frac{d^2 \vec{v}}{d t^2}.
\ee

This is the non-relativistic form of the LAD equation  which is nothing but the \textbf{Abraham-Lorentz equation }, where:
\begin{itemize}
	\item The first term \(\vec{F}_{\text{ext}}\) is the external force,
	\item The second term accounts for the radiation reaction.
	\item LAD equation  is the relativistic generalization of the Abraham-Lorentz equation  of motion, and hence have the same limitations of Runaway/Pre-Acceleration solutions. 
\end{itemize}
\subsection*{Key Challenges}

\begin{enumerate}
	\item \textbf{Runaway Solutions:}
	The LAD equation  permits solutions where the particle's acceleration grows exponentially even in the absence of external forces, which is unphysical.
	
	\item \textbf{Preacceleration:}
	The equation  predicts that a particle may start accelerating before the external force is applied, violating causality.
	
	\item \textbf{Third-Order Nature:}
	The presence of a third time derivative (\(\frac{d^3 x}{d t^3}\)) makes the LAD equation  challenging to interpret as a standard Newtonian equation  of motion.
\end{enumerate}

\subsection{Resolution: Landau-Lifshitz Approximation}

To address the problems of the LAD equation , the \textbf{Landau-Lifshitz equation } is often used. It approximates the radiation reaction force by treating it perturbatively. This removes the runaway and preacceleration problems while maintaining agreement with the LAD equation  in physical scenarios. They realized that if the second (RR) term   were much smaller than the first in the instantaneous rest frame of the charge, it would be possible to reduce the order of the LAD equation  by substituting $\frac{du}{d\tau} \to qF^{\mu\nu}u_\nu / m$ in the RR term. The result, called the Landau–Lifshitz equation , is first-order in the electron momentum and free from the pathological solutions of the LAD equation  \cite{Burton03042014}. 

As we know the LAD equation  is given as:

\be
m \frac{d u^\mu}{d \tau} = F^\mu_{\text{ext}} + \frac{q^2}{6 \pi \epsilon_0 c^3} \left( \frac{d^2 u^\mu}{d \tau^2} + \frac{u^\mu}{c^2} \frac{d u_\nu}{d \tau} \frac{d u^\nu}{d \tau} \right),
\ee

\be
m \frac{d u^\mu}{d \tau} = F^\mu_{\text{ext}} + \frac{q^2}{6 \pi \epsilon_0 c^3} \frac{d^2 u^\mu}{d \tau^2} + \frac{u^\mu}{c^2} \boxed{\frac{q^2}{6 \pi \epsilon_0 c^3} \frac{d u_\nu}{d \tau} \frac{d u^\nu}{d \tau}},
\ee

The term in the box is actually the Power radiated or in another word the amount of the energy lost and hence can be written as:
$$ \frac{d\mathcal{E}}{dt} =  - \frac{q^2}{6 \pi \epsilon_0 c^3} a^\mu a_\mu$$
where, $a^\mu = du^\mu/d\tau$ is the four-acceleration. So the term in the box makes sense as the main damping term.

\subsubsection{Classical Electron Radius and Compton Wavelength}
The classical electron radius \( r_e \) is a quantity that characterizes the effective size of the electron in classical electrodynamics, derived by equating the electron’s electrostatic potential energy with its rest energy. It represents the scale at which classical electrodynamics reaches its limits for describing the electron, as it assumes a point particle with an associated electric field.

The classical electron radius \( r_e \) is derived by equating the electron’s electrostatic potential energy to its rest energy. The rest energy of an electron is given by:
\be
E_{\text{rest}} = m_e c^2,
\ee
where \( m_e \) is the electron's rest mass and \( c \) is the speed of light. The electrostatic potential energy of a uniformly charged sphere with radius \( r_e \) and charge \( e \) is:
\be
U_{\text{electrostatic}} = \frac{1}{4 \pi \epsilon_0} \frac{e^2}{r_e}.
\ee
Equating these two energies gives:
\be
m_e c^2 = \frac{1}{4 \pi \epsilon_0} \frac{e^2}{r_e}.
\ee
Solving for \( r_e \), we get:
\be
r_e = \frac{e^2}{4 \pi \epsilon_0 m_e c^2}.
\ee
Substituting known constants:
\be
r_e \approx 2.817 \times 10^{-15} \, \text{m}.
\ee
The classical electron radius represents a characteristic length scale for the electron’s electric field, and is used in classical theories like Thomson scattering.  

The Compton wavelength \( \lambda_C \) is a quantum mechanical property of a particle. It is the wavelength of a photon whose energy is equal to the rest energy of the particle. The Compton wavelength gives a scale below which quantum effects become significant in describing a particle:
\be
\lambda_C = \frac{\hbar}{m_e c} \approx 3.8616 \times 10^{-13} \, \text{m}.
\ee
 
   The Compton wavelength \( \lambda_C \) is about 2 order larger than the classical electron radius \( r_e \):
   \be
   \frac{\lambda_C}{r_e} \approx \frac{3.8616 \times 10^{-13} }{2.817 \times 10^{-15}} \approx 137 = \frac{1}{\alpha}.
   \ee

\subsubsection{Classical and Quantum Critical fields} 
 
 The \textbf{Classical Critical Field} refers to the electromagnetic field strength at which the energy of the electromagnetic field acting on a charged particle becomes comparable to the rest energy of the particle. This is a critical threshold in classical electrodynamics, indicating when the particle can no longer be treated as a non-relativistic point particle without significant relativistic effects.

The critical field is given by the expression:

\be
E_{\text{crit}} = \frac{m_e c^2}{e r_e}
\ee

\textbf{Interpretation}:
\begin{itemize}
    \item \textbf{Field Strength}: The critical field \( E_{\text{crit}} \) represents the strength of an external electric field at which the energy associated with the field acting on a particle becomes comparable to the rest energy of the particle itself. This implies that the particle would begin to exhibit significant relativistic or quantum mechanical behavior if subjected to such a field.
    \item \textbf{Physical Meaning}: If a particle with charge \( e \) (such as an electron) is placed in an electric field of strength \( E_{\text{crit}} \), the energy gained by the particle due to the field is equal to the rest energy of the particle. This is the point at which the effects of the field become comparable to the particle's intrinsic energy, and classical treatment breaks down.
\end{itemize}

Substituting the classical electron radius \( r_e \approx 2.817 \times 10^{-15} \, \text{m} \) into the expression for \( E_{\text{crit}} \) gives the \textbf{Classical critical field} value:

\be
E_{\text{crit}} = \frac{m_e c^2}{e r_e}  \approx \, 1.813\times 10^{20}\ \text{V/m} \approx 1.813 \times 10^{18} \ \text{V/cm} .
\ee

Similary, the\textbf{ Quantum Critical field} is defined as:

\be
E_{\text{qm,crit}} = \frac{m_e c^2}{e \lambda_C} \approx  1.32 \times 10^{18}\ \text{V/m} \approx  1.32 \times 10^{16}\ \text{V/cm}
\ee
where, $\lambda_C = \hbar/m_e c$ is the Compton wavelength.  

\subsubsection{Argument by Landau-Lifshitz}
 
 The Landau and Lifshitz showed that if the electric field experieced by the electron in it's instanteneous rest frame is smaller than the Classical Critical field, i.e. $|E_{IRF}| \ll E_{crit}$ then in such scenarios the Radiation Reaction can be considered as the small perturbation on the zeroth order term, implying:
 $$ \frac{d u^\mu}{ds} \approx \frac{F^\mu_{\text{ext}}}{m_e}$$
 
So we again rewrite the LAD equation  as: 

\be
m_e \frac{d u^\mu}{ds} = F^\mu_{\text{ext}} + m_e \tau \left( \frac{d^2 u^\mu}{ds^2} + \frac{u^\mu}{c^2} \frac{d u_\nu}{ds} \frac{d u^\nu}{ds} \right),
\ee
and using the LL approximation it can be written as:

\be 
m_e \frac{d u^\mu}{ds} = F^\mu_{\text{ext}} + \tau \frac{dF^\mu_{\text{ext}} }{ds} + \mathcal{O}(\tau^2)
\ee

Here, $\tau = q^2/(6 \pi \epsilon_0 c^3 m_e) \approx 6.26\times 10^{-24}\ \text{sec}$ is the characteristic time and $s$ is the proper time. One can ignore the $\mathcal{O}(\tau^2)$ as this term will be even smaller that the quantum-corrections. Anyways the Classical critical field is itself smaller than 2 order of magnitude from the quantum critical field, so in the purview of the classical ED we are anyway ignoring the terms which are way below the quantum limit. In this sense the LL equation  is not the approximation but rather the classically exact solution for the radiation reaction. 

So if we substitute $F^\mu_{ext} = qF^{\mu\nu} u_\nu$ for the Lorentz force responsible for the motion of the electron in the intense electromagnetic field, then eventually the LL equation  will be then given by after neglecting $\mathcal{O}(\tau^2)$ as:

\be
m_e \frac{d u^\mu}{d \tau} = q F^{\mu \nu} u_\nu +  \tau \left[ \frac{q}{m_e} (\partial_\alpha F^{\mu \nu}) u^\alpha u_\nu - \frac{q^2}{m_e^2} F^{\mu \nu} F_{\alpha \nu} u^\alpha + \frac{q^2}{m_e^2 c^2} (F^{\alpha \nu} u_\nu)(F_{\alpha \lambda} u^\lambda) u^\mu \right]
\ee

It should be noted that the though the LL equation  is an approximation but the difference between LAD and LL equations are smaller than the quantum effects which we have already neglected. In that sense classicaly it is an exact equation  of motion which takes into account the Radiation Reaction without any issues of ``Runaway'' solutions, i.e. in the absence of the external force all the RHS becomes zero and we have just a particle moving with constant velocity. 

\subsection{Litrature on Radiation Reaction}

The field of Radiation Reaction (RR) is very rich and remains an area of active research. There are several elaborate review articles, such as \cite{Burton03042014, Blackburn2020}. Additionally, there are various theoretical models of RR, mostly based on one or another variant of the Landau-Lifshitz equations \cite{Sokolov2009, Piazza2008, FORD1993182, PhysRevAccelBeams.24.114002, Tamburini_2010}, among many others. Although the list of references is not exhaustive, readers are encouraged to appreciate this vast field of Radiation Reaction from both the Classical and Quantum perspectives 

\section{Dimensionless Units}
We will be now looking at the numerical examples and for that it is better to define dimensionless units. 

\subsection{Normalized variables}
The typical Lorentz force equation  is written as: 

$$m \frac{d^2 \mathbf{r}}{dt^2} = q (\mathbf{E} + \mathbf{v} \times \mathbf{B})$$

We can make this equation  dimensionless as:
\begin{itemize}
    \item $\beta = v/c$ : velocity 
    \item $m' = m/m_e$ : mass
    \item $q' = q/e$ : charge
    \item $r' = k r$ : position 
    \item $t' = \omega t$ : time ($\omega = kc$ is frequency of laser)
    \item $p' = p/m_ec$ : momentum
    \item $a_0 = eE_0/m_e\omega c = eB_0/m_e\omega$ : dimensionless electric and magnetic field amplitude
    \item $\varphi = e\phi/m_ec^2$ : scalar potential
    \item $a_0 = eA_0/m_e c$ : vector potential 
\end{itemize}

\subsection{Intensity normalization}

The intensity \( I \) is related to the peak electric field \( E_{\text{peak}} \) through the Poynting vector:

\be
\mathbf{S} = \frac{1}{\mu_0} \mathbf{E} \times \mathbf{B}
\ee

Since \( |\mathbf{B}| = |\mathbf{E}|/c \), we have:

\be
|\mathbf{S}| = \frac{1}{\mu_0 c} E_{\text{peak}}^2
\ee

The time-averaged intensity is:

\be
I_0 = \langle I \rangle = \frac{1}{2} \cdot \frac{1}{\mu_0 c} E_{\text{peak}}^2
\ee

\be
I_0 = \langle I \rangle = \frac{1}{2} \cdot \frac{1}{\mu_0 c} \left( \frac{a_0 m_e c \omega}{e} \right)^2
\ee

Now, using \( \omega = \frac{2\pi c}{\lambda} \):

\be
I_0 = \langle I \rangle = \frac{1}{2} \cdot \frac{1}{\mu_0 c} \left( \frac{a_0 m_e c \frac{2\pi c}{\lambda}}{e} \right)^2
\ee

Simplifying:

\be
I_0 = \frac{a_0^2 m_e^2 c^3 (2\pi)^2}{2 e^2 \mu_0 \lambda^2} \implies
I_0 \lambda^2 = \Big(\frac{m_e^2 c^3 (2\pi)^2}{2 e^2 \mu_0}\Big) a_0^2 \ \quad \text{W/m}^2\  \text{m}^2
\ee

\be
I_0 \lambda^2 = 10^{-4} 10^{12} \Big(\frac{m_e^2 c^3 (2\pi)^2}{2 e^2 \mu_0}\Big) a_0^2 \quad \ \text{W/cm}^2\  \mu\text{m}^2 \approx 1.3681671 \times 10^{18}\ \  a_0^2 \quad \ \text{W/cm}^2\  \mu\text{m}^2
\ee

If \( a_0 > 1 \), relativistic dynamics will play an important role. Working with dimensionless units is very useful; not only can we get rid of numerical errors during calculations, but the problem also becomes scalable. For example, if the wavelength of the laser is varied, the main characteristics of the problem will scale accordingly.

\section{Fig 8 Trajectory Problem}
Now let us consider the interaction of an electron with a linearly polarized plane pulse, whose vector potential is  given by:
$$\mathbf{A} = a_0 \cos(\phi)~~ \hat{x}$$
where, $\phi = \omega t - kz$ is the phase. We have not considered any field envelope at this moment to keep the derivation simple to follow. 

The corresponding electric and magnetic fields of the laser pulse can be obtained to be:

$$\mathbf{E} = -\frac{\partial \mathbf{A}}{\partial t} = a_0 \omega \sin(\phi)~~\hat{x} \quad;\quad
 \mathbf{B} = \nabla \times \mathbf{A} = a_0 k \sin(\phi)~~ \hat{y} $$

The total energy of the electron is given as $\mathcal{E} = \gamma m_e c^2$ and from the relativistic Lorentz force equation  we know that :
$$ \frac{d\mathcal{E}}{dt} = q \left( \mathbf{E} \cdot \mathbf{v} \right) \implies 
\frac{d\mathcal{\gamma}}{dt} = \frac{q}{m_ec^2} a_0 \omega \sin(\phi) v_x 
$$
Furthermore, it can be readily be found that:

$$ 
\frac{dp_x}{dt} = q [E_x - B_y v_z] = q [1 - v_z/c] a_0 \omega \sin(\phi) \quad ; \quad 
\frac{dp_y}{dt} = 0 \quad ; \quad 
\frac{dp_z}{dt} = q B_y v_x = q a_0 \omega \sin(\phi) \frac{v_x}{c}
$$

It can be seen that: 
$$ 
\frac{d\mathcal{\gamma}}{dt} = \frac{1}{m_e c} \frac{dp_z}{dt} \implies \frac{d}{dt}\Big(\gamma - \frac{p_z}{m_e c} \Big) = 0 \implies \boxed{\gamma - \frac{p_z}{m_ec} = \alpha\quad \text{(some constant)} }
$$
Using the fact $\gamma = \sqrt{1 + \mathbf{p}\cdot\mathbf{p} /m_e^2c^2} $,  it can be shown that:

$$\frac{p_z}{m_ec} = \frac{1}{2\alpha} [1 - \alpha^2 + p_\perp^2/m_e^2c^2] $$ 
where, $p_\perp = \sqrt{p_x^2+p_y^2}$. 

For the initial condition at $t = 0$ we have $p_x = p_y = p_z = 0$ which implies $\gamma(t = 0) = 1$ 
and hence $\alpha = 1$ for this case. The above equation  then simplifies for $p_y = 0$ as for linearly
polarized it wont change: 
$$ \frac{p_z}{m_e c} = \frac{p_x^2}{2m_e^2c^2}$$

\subsection*{Deriving \( p_z(\phi) \)}
We now proceed to derive \( p_z(\phi) \). The equation  for \( \frac{dp_z}{dt} \) is:

\be
\frac{dp_z}{dt} = q a_0 \omega \sin(\phi) \frac{v_x}{c}
\ee

Using the chain rule:

\be
\frac{dp_z}{dt} = \frac{dp_z}{d\phi} \cdot \frac{d\phi}{dt}
\ee

The derivative of \( \phi = \omega t - kz \) is:

\be
\frac{d\phi}{dt} = \omega - \frac{\omega v_z}{c} = \omega \left( 1 - \frac{v_z}{c} \right)
\ee

As we have $\gamma - p_z/m_ec = 1 \implies 1 - v_z/c = 1/\gamma $:

\be
\frac{d\phi}{dt} = \frac{\omega}{\gamma} \label{2}
\ee

Substituting this into the equation  for \( \frac{dp_z}{dt} \):

\be
\frac{dp_z}{d\phi} = q a_0 \sin(\phi) \sqrt{\frac{2p_z}{m_e c}} \label{3}
\ee

Integrating gives:

\be
p_z(\phi) = \frac{q^2 a_0^2}{4 m_e c} \left( 1 + \cos(2\phi) \right) \label{4}
\ee

\subsection*{Deriving \( p_x(\phi) \)}

From the equation  for \( \frac{dp_x}{dt} \), we have with $\gamma - p_z/m_ec = 1 \implies 1 - v_z/c = 1/\gamma$:

\be
\frac{dp_x}{dt} = q a_0 \omega \sin(\phi)  \frac{1}{\gamma} \label{5}
\ee

Using the chain rule:

\be
\frac{dp_x}{d\phi} \cdot \frac{\omega}{\gamma} = \frac{1}{\gamma}q a_0 \omega \sin(\phi)  
\ee

Simplifying:

\be
\frac{dp_x}{d\phi} = q a_0 \sin(\phi)   \label{6}
\ee
 
Integrating:

\be
p_x(\phi) = -q a_0 \cos(\phi) 
\ee

\subsection*{Deriving \( x(\phi) \) and $z(\phi)$}

To obtain the position components \( x(\phi) \) and \( z(\phi) \) from the momentum components \( p_x(\phi) \) and \( p_z(\phi) \), we use the following relations from relativistic mechanics:

\be
p_x = \gamma m_e \frac{dx}{dt}, \quad p_z = \gamma m_e \frac{dz}{dt}
\ee

where \( p_x(\phi) \) and \( p_z(\phi) \) are the momentum components as a function of the phase \( \phi \), and \( \gamma \) is the Lorentz factor given by:

\be
\gamma = \frac{1}{\sqrt{1 - v_x^2/c^2 - v_z^2/c^2}}
\ee

First, we express the velocities in terms of the momenta. From the above relations:

\be
\frac{dx}{dt} = \frac{p_x}{\gamma m_e}, \quad \frac{dz}{dt} = \frac{p_z}{\gamma m_e}
\ee

Thus, we need to integrate \( \frac{dx}{dt} \) and \( \frac{dz}{dt} \) with respect to \( \phi \).
 
Using the chain rule:

\be
\frac{d\phi}{dt} = \omega \left( 1 - \frac{v_z}{c} \right) = \frac{\omega}{\gamma}
\ee

So:

\be
\frac{dt}{d\phi} = \frac{\gamma}{\omega}
\ee

This will allow us to convert the time derivatives into phase derivatives.
 
Using the fact that \( \frac{dx}{dt} = \frac{p_x}{\gamma m_e} \) and \( \frac{dz}{dt} = \frac{p_z}{\gamma m_e} \), we can write:

\be
\frac{dx}{d\phi} = \frac{p_x}{\gamma m_e} \cdot \frac{dt}{d\phi} = \frac{p_x}{\gamma m_e} \cdot \frac{\gamma}{\omega} = \frac{p_x}{m_e \omega}
\ee

\be
\frac{dz}{d\phi} = \frac{p_z}{\gamma m_e} \cdot \frac{dt}{d\phi} = \frac{p_z}{\gamma m_e} \cdot \frac{\gamma}{\omega} = \frac{p_z}{m_e \omega}
\ee
 
Now we integrate to find the position as a function of the phase \( \phi \).

For \( x(\phi) \):

\be
x(\phi) = \int \frac{p_x(\phi)}{m_e \omega} d\phi
\ee

Substitute the expression for \( p_x(\phi) \):

\be
x(\phi) = \int \frac{-q a_0 \cos(\phi)}{m_e \omega} d\phi
\ee

This gives:

\be
x(\phi) = \frac{q a_0}{m_e \omega} \sin(\phi)
\ee

For \( z(\phi) \):

\be
z(\phi) = \int \frac{p_z(\phi)}{m_e \omega} d\phi
\ee

Substitute the expression for \( p_z(\phi) \):

\be
z(\phi) = \int \frac{q^2 a_0^2}{4 m_e^2 c \omega} \left( 1 + \cos(2\phi) \right) d\phi
\ee

This gives:

\be
z(\phi) = \frac{q^2 a_0^2}{4 m_e^2 c \omega} \left( \phi + \frac{1}{2} \sin(2\phi) \right)
\ee

Thus, the positions \( x(\phi) \) and \( z(\phi) \) as a function of \( \phi \) are:

\be
x(\phi) = \frac{q a_0}{m_e \omega} \sin(\phi)
\ee

\be
z(\phi) = \frac{q^2 a_0^2}{4 m_e^2 c \omega} \left( \phi + \frac{1}{2} \sin(2\phi) \right)
\ee

If we consider $a_0 \leftarrow ea_0/m_e\omega$ and $z \leftarrow k z$ then we have the dimensionless form
of the above equations as

\be
x(\phi) = a_0 \sin(\phi)
\ee

\be
z(\phi) = \frac{a_0^2}{4} \left( \phi + \frac{1}{2} \sin(2\phi) \right)
\ee

\subsection{Summary of the equations obtained in dimensionless form}

\be
p_z(\phi) = \frac{a_0^2}{4} \left( 1 + \cos(2\phi) \right) \label{4} \implies \langle p_z \rangle = \frac{a_0^2}{4}
\ee
here, we also estiamted the cycle average. $\langle \cos(2\phi)\rangle = 0$ as it is very rapidly oscillating function. 

\be
z(\phi) = \frac{a_0^2}{4} \left( \phi + \frac{1}{2} \sin(2\phi) \right)
\ee

\be
p_x(\phi) = -a_0 \cos(\phi) \implies  \langle p_x^2 \rangle = \frac{a_0^2}{2}
\ee

\be
x(\phi) = a_0 \sin(\phi)
\ee
\subsection{Trajectory of particle in average rest frame}

We also obtained earlier,

$$\frac{p_z}{m_ec} = \frac{1}{2\alpha} [1 - \alpha^2 + p_\perp^2/m_e^2c^2] $$ 
which in dimensionless form becomes with $p_y = 0$

$$p_z = \frac{1}{2\alpha} [1 - \alpha^2 + p_x^2] \implies 
\langle p_z \rangle = \frac{1}{2\alpha} [1 - \alpha^2 + \langle p_x^2 \rangle] = \frac{1}{2\alpha} [1 - \alpha^2 + \frac{a_0^2}{2}] 
$$ 

Therefore, for the average rest frame $\langle p_z \rangle = 0 \implies \boxed{\alpha = \sqrt{1 + a_0^2/2}}$. 
Substituting the value of $\alpha$ and $p_x^2 = a_0^2 \cos^2(\phi)$ we will have:
$$ p_z(\phi) = \frac{a_0^2}{4\sqrt{1+a_0^2/2}} \cos(2\phi) \quad;\quad p_x = -a_0\cos(\phi)$$

\begin{figure}[!t]
    \centering
    \includegraphics[width=0.5\textwidth]{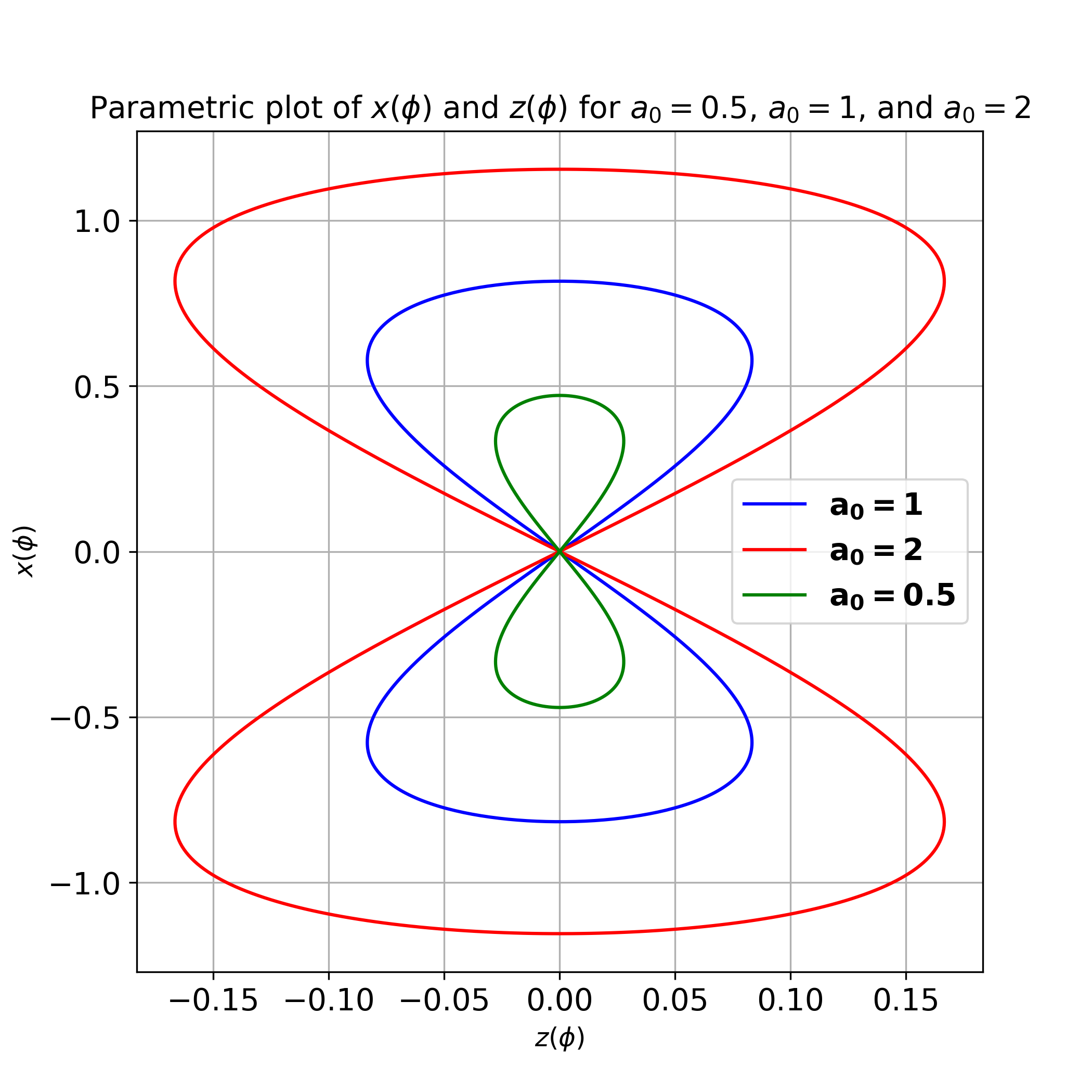}
    \caption{The parametric plot in the $x$ and $z$ coordinate is shown for the IRF of the particle. The laser is polarized along the $x$ direction. }
    \label{fig2}
\end{figure}

Now, 
$$\mb{p} = \gamma m_e \frac{d\mb{r}}{dt} = \gamma m_e \frac{d\mb{r}}{d\phi} \frac{d\phi}{dt} = \gamma m_e \frac{d\mb{r}}{d\phi} \alpha\frac{\omega}{\gamma} $$

In dimensionless form,
$$ \mb{p} = \alpha \frac{d\mb{r}}{d\phi} \implies 
x(\phi) = \frac{1}{\alpha}\int p_x d\phi \ ; \ z(\phi) = \frac{1}{\alpha}\int p_z d\phi $$

therefore, finally we obtain the trajectory of the particle in the average rest frame of the particle and given as:

$$ \boxed{x(\phi) = \frac{a_0}{\sqrt{1+a_0^2/2}} \sin(\phi)\quad;\quad z(\phi) =  \frac{a_0^2}{8(1+a_0^2/2)} \sin(2\phi) } $$

The parametric plot of the particle trajectory in IRF is shown in Fig. \ref{fig2}.

\section{Canonical Momenta and Constants of motion}

Given the Lagrangian for a relativistic charged particle in an electromagnetic field:
\be
\mathcal{L} = -m c^2 \sqrt{1 - \frac{\mathbf{v}^2}{c^2}} + q \mathbf{v} \cdot \mathbf{A} - q \phi,
\ee
the canonical momentum $\mathbf{p}$ is defined as:
\be
\mathbf{p} = \frac{\partial \mathcal{L}}{\partial \mathbf{v}}.
\ee

\subsection{Relativistic Kinetic Term}
The derivative of the relativistic kinetic term is:
\be
\frac{\partial}{\partial \mathbf{v}} \left( -m c^2 \sqrt{1 - \frac{\mathbf{v}^2}{c^2}} \right) = \frac{m \mathbf{v}}{\sqrt{1 - \frac{\mathbf{v}^2}{c^2}}},
\ee
which is the relativistic mechanical momentum.

\subsection{Interaction with the Vector Potential}
The derivative of the term involving the vector potential is:
\be
\frac{\partial}{\partial \mathbf{v}} \left( q \mathbf{v} \cdot \mathbf{A} \right) = q \mathbf{A},
\ee
since the vector potential $\mathbf{A}$ does not depend on $\mathbf{v}$.

\subsection{Electric Scalar Potential}
The scalar potential term $-q \phi$ does not depend on $\mathbf{v}$, so its derivative is:
\be
\frac{\partial}{\partial \mathbf{v}} \left( -q \phi \right) = 0.
\ee

\subsection{Final Expression}
Combining these results, the canonical momentum is:
\be
\mathbf{p} = \frac{m \mathbf{v}}{\sqrt{1 - \frac{\mathbf{v}^2}{c^2}}} + q \mathbf{A} = \gamma m \mathbf{v} + q \mathbf{A}.
\ee

\subsection{Key Points}
\begin{itemize}
    \item The term $\frac{m \mathbf{v}}{\sqrt{1 - \frac{\mathbf{v}^2}{c^2}}}$ is the \textbf{relativistic mechanical momentum}.
    \item The term $q \mathbf{A}$ is the contribution due to the interaction with the electromagnetic field, arising from the vector potential.
\end{itemize}

This distinction between the canonical momentum and the mechanical momentum is essential in the Hamiltonian formulation and quantum mechanics.	

\subsection{Cyclic Coordinates and Conservation Laws}

In Lagrangian mechanics, a \textbf{cyclic coordinate} (or \textbf{ignorable coordinate}) is a generalized coordinate that does not explicitly appear in the Lagrangian. If a coordinate $q_i$ is cyclic, the corresponding conjugate momentum $p_i$ is conserved.

\subsection{Definition of Cyclic Coordinates}

Consider a Lagrangian $\mathcal{L}(q_1, q_2, \dots, q_n; \dot{q}_1, \dot{q}_2, \dots, \dot{q}_n; t)$, where $q_i$ are the generalized coordinates and $\dot{q}_i$ are the corresponding velocities. A coordinate $q_i$ is said to be cyclic if:
\be
\frac{\partial \mathcal{L}}{\partial q_i} = 0.
\ee

This means the Lagrangian does not explicitly depend on $q_i$.

\subsection{Conservation of Conjugate Momentum}

For any coordinate $q_i$, the conjugate momentum is defined as:
\be
p_i = \frac{\partial \mathcal{L}}{\partial \dot{q}_i}.
\ee

If $q_i$ is cyclic, the Euler-Lagrange equation :
\be
\frac{d}{dt} \frac{\partial \mathcal{L}}{\partial \dot{q}_i} - \frac{\partial \mathcal{L}}{\partial q_i} = 0
\ee
reduces to:
\be
\frac{d}{dt} \left( \frac{\partial \mathcal{L}}{\partial \dot{q}_i} \right) = 0.
\ee

Thus, the conjugate momentum $p_i$ is conserved:
\be
p_i = \text{constant}.
\ee

\subsection{Physical Significance}

\begin{itemize}
    \item A cyclic coordinate represents a symmetry in the system.
    \item The conservation of the conjugate momentum $p_i$ is a direct consequence of Noether's theorem, which relates symmetries to conservation laws.
    \item For example:
    \begin{itemize}
        \item If the Lagrangian does not depend on a coordinate representing position, the conjugate momentum corresponds to linear momentum, which is conserved.
        \item If the Lagrangian does not depend on an angular coordinate, the conjugate momentum corresponds to angular momentum, which is conserved.
    \end{itemize}
\end{itemize}

\section{Dynamics in Gaussian (temporal only) Laser profile} 

\subsection{Basics of circularly polarized laser-electron interaction}

For a circularly polarized light $E_z = B_z = 0$, in such case simplify the Lorentz force equation . 
 
 The relativistic Lorentz force law in 4-vector form is given by:
\be
\frac{dp^\mu}{d\tau} = q F^{\mu \nu} u_\nu,
\ee
where:
\begin{itemize}
    \item \( p^\mu \) is the 4-momentum of the particle,
    \item \( F^{\mu \nu} \) is the electromagnetic field tensor,
    \item \( u_\nu \) is the 4-velocity of the particle,
    \item \( q \) is the charge of the particle,
    \item \( \tau \) is the proper time.
\end{itemize}

The 4-momentum \( p^\mu \) of a particle is defined as:
\be
p^\mu = m u^\mu = m \gamma (c, v_x, v_y, v_z),
\ee
where:
\begin{itemize}
    \item \( m \) is the rest mass of the particle,
    \item \( \gamma = \frac{1}{\sqrt{1 - v^2/c^2}} \) is the Lorentz factor,
    \item \( (v_x, v_y, v_z) \) are the components of the particle’s velocity.
\end{itemize}

The electromagnetic field tensor \( F^{\mu \nu} \) in SI units is:
\be
F^{\mu \nu} = 
\begin{pmatrix} 
0 & -E_x/c & -E_y/c & 0 \\
E_x/c & 0 & -B_z & B_y \\
E_y/c & B_z & 0 & -B_x \\
0 & -B_y & B_x & 0
\end{pmatrix},
\ee
where \( E_x, E_y, E_z \) are the components of the electric field and \( B_x, B_y, B_z \) are the components of the magnetic field.

Now, we compute the time and spatial components of the Lorentz force law under the condition \( E_z = 0 \) and \( B_z = 0 \).
 
The time component of the equation  is:
\be
\frac{dp^0}{d\tau} = q \sum_{\nu=0}^{3} F^{0\nu} u_\nu.
\ee
Substituting the relevant components and using \( E_z = 0 \), \( B_z = 0 \), we get:
\be
\frac{dp^0}{d\tau} = \frac{q\gamma}{c} (E_x v_x + E_y v_y).
\ee

The spatial z-component of the equation  is:
\be
\frac{dp^3}{d\tau} = q \sum_{\nu=0}^{3} F^{3\nu} u_\nu.
\ee
After substituting the relevant components, we get:
\be
\frac{dp^3}{d\tau} = q \gamma (v_x B_y - v_y B_x).
\ee

To express the equations in terms of the coordinate time derivative \( \frac{d}{dt} \), we use the relation:
\be
\frac{d}{dt} = \gamma \frac{d}{d\tau}.
\ee
Thus, we have:

\be
\frac{dp^0}{dt} = \frac{q}{c} (E_x v_x + E_y v_y).
\ee

\be
\frac{dp^3}{dt} = q (v_x B_y - v_y B_x).
\ee
 
\begin{itemize}
    \item Time Component:
    \be
    \frac{dp^0}{dt} = \frac{q}{c} (E_x v_x + E_y v_y). \label{dp0dt}
    \ee
    \item Spatial z-Component:
    \be
    \frac{dp^3}{dt} = q (v_x B_y - v_y B_x).\label{dp3dt}
    \ee
\end{itemize}

These equations describe the relativistic Lorentz force law in terms of the coordinate time derivative, with the electric and magnetic field components restricted to the \( x \)-\( y \) plane (\( E_z = 0 \) and \( B_z = 0 \)).

 \subsection{Motion in Gaussian circularly polarized  plane wave}
We consider the interaction of the relativistic electron(s) with the counter-propagating Gaussian laser pulse. The direction of the laser propagation is considered to be along the $z$ axis. The plane laser pulse having a frequency $\omega$ and a FWHM pulse duration of $\tau_0$ is expressed in the following vector potential form,
\be  \mb{A} = a_0\ \exp\left(- \frac{\alpha\eta^2}{\tau_0^2}\right) \Bigl[\delta \cos(\eta) \mb{e_x} + \sqrt{1 -\delta^2} \sin(\eta)\ \mb{e_y} \Bigr]  \label{vecPot} \ee 
where, $a_0 = eA_0/m_ec$ is the dimensionless amplitude of the vector potential, $\eta = \omega t - k z + \phi_0$, $\phi_0$ is some phase constant, and $k$  is the wave number of the laser pulse. Furthermore, $a_0$ denotes the magnitude of the dimensionless vector potential, $\delta$ is a parameter which can be 1 ($1/\sqrt{2}$) for linearly (circularly) polarized light respectively, and $\alpha \equiv 4 \ln(2)$ for Gaussian laser pulse.

As can be seen from equation \eqref{vecPot} that the vector potential is not a function of $x$ and $y$ and hence corresponding conjugate momenta $P_x$ and $P_y$ will be ca constant of motion. We also know that for the charge particle interacting with the elctromagnetic fields the conjugate momenta or canonical momenta is given by:
\be \mb{P_c} = \gamma m \mb{v} + q \mb{A} \equiv \mb{P} + q \mb{A}  \ee
where, $\mb{P} = \gamma m \mb{v}$ represents the kinetic momenta and $\mb{P_c}$ is the corresponding canonical momenta. As we are dealing with electron and hence in the dimensionless units $q = -1$ and with $\mb{P_c}_\perp$ (perpendicular to direction of propagation of electromagnetic waves) being the constant of the motion (as the Lagrangian does not depend on the $x$ and $y$ coordinates and space dependence only come in the form of $\eta = t -z$), which leads to $\mb{P}_\perp - \mb{A}_\perp = \text{Constant}$. 

We will be now using the dimensionless units. The dimensionless space and time are taken in units of the wave number $k$ and the angular frequency $\omega$ respectively as discussed earlier.  Laser electric field is normalized as $\mb{a} = |e| \mb{E} / m_e \omega c$  with $e$ and $m_e$ being the charge and the mass of the electron. By using  the dimensionless form of the problem, $\eta$ can be written as $\eta = t - z + \phi_0$.  The definition of the electric field $\mb{E} = - \partial \mb{A}/\partial t$ and magnetic field $\mb{B} = \nabla \times \mb{A}$,  leads to the following dimensionless form  of the fields:
\begin{align}
E_x &= \delta a_0 \exp\left(- \frac{\alpha\eta^2}{\tau_0^2}\right)  \Bigl[\sin(\eta) + \frac{2\alpha\eta}{\tau_0^2} \cos(\eta)\Bigr], \label{ex} \\  
E_y &= \sqrt{1 -\delta^2}\ a_0 \exp\left(- \frac{\alpha\eta^2}{\tau_0^2}\right) \Bigl[\frac{2\alpha\eta}{\tau_0^2} \sin(\eta) - \cos(\eta)\Bigr], \label{ey}\\
E_z &= 0, \label{ez}\\
B_x &= -\sqrt{1 -\delta^2}\ a_0 \exp\left(- \frac{\alpha\eta^2}{\tau_0^2}\right) \Bigl[\frac{2\alpha\eta}{\tau_0^2} \sin (\eta) - \cos(\eta)\Bigr], \label{bx}\\
B_y &= \delta a_0 \exp\left(- \frac{\alpha\eta^2}{\tau_0^2}\right)  \Bigl[\sin(\eta) + \frac{2\alpha\eta}{\tau_0^2} \cos(\eta)\Bigr], \label{by}\\
B_z &= 0 \label{bz}
\end{align}

The dynamics of the charge particle is governed by the Lorentz force is given by, 
\be \frac{dP^\mu}{d\tau} = - F^{\mu\nu} u_\nu  \label{lorentz}\ee
where, $\tau$ is the proper time, $P^{\mu} = u^\mu$ is the four momentum, $F^{\mu\nu} = \partial^\mu A^\nu - \partial^\nu A^\mu$ is electromagnetic field tensor and $A^\mu = (0,\mb{A})$ is four potential corresponding to laser. The four velocity of the particle(s) is denoted by $u^\mu = (\gamma ,\gamma \mb{v})$, where the relativistic factor $\gamma = 1/\sqrt{1 - \mb{v}^2}$.  
From equation \ref{lorentz} and Eqs. \ref{ex}-\ref{bz}, it can be shown that the rate of the change of energy of the particle in dimensionless form is given by ,
\be\frac{d\gamma}{dt} =  \frac{dP_z}{dt}. \label{dgdt}\ee
This is obtained by noting that $E_x = B_y$ and $E_y = -B_x$, use this in Eqs. \eqref{dp0dt} and \eqref{dp3dt} and the equation \eqref{dgdt} can be easily obtained. 

The above equation  can be easily  integrated to have the relation between the relativistic factor $\gamma$ and the $z$ component of the momentum  of the particle $P_z$ as,
\be \xi_z = \gamma - P_z  \label{xiz}\ee
here, $\xi_z$ is the constant of motion. Furthermore, from the $x$ and the $y$ components of the Lorentz force (equation \ref{lorentz}),  other constants of the motion can be deduced by solving the equation ,
\be \frac{d}{dt} (\mb{P}_\perp - \mb{A}_\perp) = 0\ee
this would result in the constants of the motion $\xi_\perp = \mb{P}_\perp - \mb{A}_\perp$. The initial conditions i.e. at $t = 0$, $\mb{A}_\perp = 0$ would result in $\mb{P}_\perp = \mb{A}_\perp$. It can be shown that the relation $\mb{P} = \gamma \mb{v}$ and the definition $\mb{B} = \nabla \times \mb{A}$ would result in following force along the $z$ direction.
\be\frac{dP_z}{dt} = \upsilon_y B_x - \upsilon_x B_y = - \frac{1}{\gamma} \mb{P}_\perp \cdot \frac{\partial \mb{A}_\perp}{\partial z} = -\frac{1}{2\gamma} \frac{\partial \mb{A}_\perp^2}{\partial z} \label{dpzdt}.\ee
The temporal evolution of the relativistic gamma factor can be estimated from equation \ref{dgdt} and equation \ref{dpzdt} as,
\be \gamma \frac{d\gamma}{dt} = -\frac{1}{2} \frac{\partial \mb{A}_\perp^2}{\partial z}\ee
It can be simplified if we write the same in the co-moving frame of the laser i.e. in terms of $\eta = t - z$
\be \gamma\frac{d\eta}{dt} \frac{d\gamma}{d\eta} = \frac{1}{2} \frac{\partial \mb{A}_\perp^2}{\partial \eta} \label{gaeta}\ee
Furthermore, $\eta = t - z$ leads to,
\be \frac{d\eta}{dt} = 1 - \frac{Pz}{\gamma} \implies \gamma\frac{d\eta}{dt} = \xi_z.\ee
The constant of motion $\xi_z = \gamma - P_z$ (equation \ref{xiz}) can be obtained by the initial condition that $t = 0$, $P_z = -\gamma_0 \sqrt{1 - \gamma_0^{-2}} \sim - \gamma_0$ which implies $\xi_z \sim 2\gamma_0$, and equation \ref{gaeta} can be directly integrated as:
\be \boxed{\gamma(\eta) \sim \frac{1}{4\gamma_0} \mb{A}_\perp^2(\eta) + \gamma_0 \label{geta2}}\ee

\begin{figure}[!t]
    \centering\includegraphics[width=\textwidth]{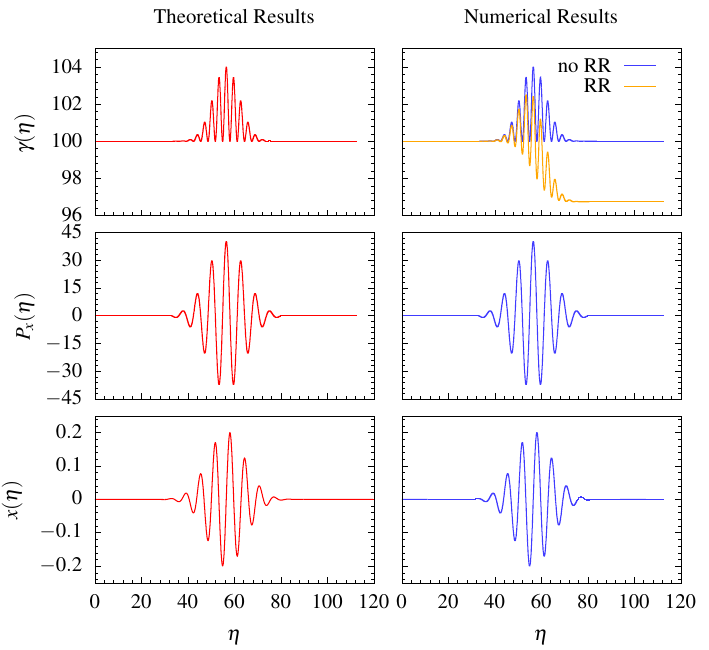}
    \caption{Comparing the analytical and numerical results for $a_0 = 40$ Gaussian, linearly polarized, laser pulse of 3 cycle interacting with counter-propagating electron having $\gamma_0 = 100$. Effect of the Radiation Reaction is also shown for $\gamma(\eta)$ in the numerical results, as the analytical solution with RR is too involved to be discussed here.}
\end{figure}

The trajectory of the particle can easily be obtained by integrating the following equations:
\be \frac{d\mb{r}}{d\eta} \frac{d\eta}{dt} = \frac{\mb{P}}{\gamma}  \ee
\be \frac{d\mb{r}}{d\eta} = \frac{\mb{P}}{\xi_z} = \frac{\mb{P}}{2\gamma_0} \label{reta} \ee
The perpendicular position can be obtained by utilizing the fact that $\boxed{\mb{P}_\perp = \mb{A}_\perp}$ as:
\be \boxed{\mb{r}_\perp(\eta) = \frac{1}{2\gamma_0} \int \mb{A}_\perp d\eta + \mb{r_0} }\ee
 here, $\mb{r_0}$ needs to be evaluated by initial conditions. 
 
Similarly $z$ component can be written as from equation \ref{reta} as:
\be \frac{dz}{d\eta} = \frac{P_z}{2\gamma_0}  \ee
From, equation \ref{dgdt} we can equate $\boxed{\gamma(\eta) = P_z(\eta) + \xi_z \sim P_z(\eta) + 2 \gamma_0}$. 
\be z(\eta) = \frac{1}{2\gamma_0} \int P_z d\eta + z_0 = \frac{1}{2\gamma_0} \int [\gamma(\eta) - 2\gamma_0] d\eta + z_0 \ee
\be z(\eta) = -\eta + \frac{1}{2\gamma_0} \int \gamma(\eta) d\eta + z_0\ee
\be z(\eta) = -\eta + \frac{1}{2\gamma_0} \int \Big[ \frac{1}{4\gamma_0} \mb{A}_\perp^2(\eta) + \gamma_0\Big] d\eta + z_0\ee 
  
\be z(\eta) =  - \frac{\eta}{2} + \frac{1}{8\gamma_0^2}\int \mb{A}_\perp^2 d\eta + z_0.\ee
\be z(\eta) = - \frac{\eta}{2}  + \frac{1}{8\gamma_0^2} \frac{a_0^2}{2}\int \exp\left(- \frac{2\alpha\eta^2}{\tau_0^2}\right) d\eta + z_0 \ee
\be \boxed{ z(\eta) =  z_0 - \frac{\eta}{2} + \frac{a_0^2 \tau_0}{16\gamma_0^2}\sqrt{\frac{\pi}{8\alpha}}\ \text{erf}\left( \frac{\eta\sqrt{2\alpha}}{\tau_0} \right)}\ee
The value of the $z_0$ can be obtained by knowing the initial conditions.

\section{Ponderomotive scattering by paraxial laser beams}

So far we have used a plane wave without any temporal envelope and study how the figure 8 shaped trajectory emerge in the instantenous rest frame of the particle. Next, we incorporated the gaussian temporal envelope and solve the problem analytically. The results are then compared with the numerical results where the effect of the RR is also studied. Now let us consider a scenario when the laser is modelled more realistically using the paraxial beam approximation, in such case we will see how the Ponderomotive scattering angle can be calculated analytically and how this process can be used as a potential diagnostic tool to measure the physical parameters associated with the ultra-intense lasers and also as a test tool to validate the theoritical framework of Radiation Reaction. This section is based on our previous work \cite{Holkundkar_2022}. 

\subsection{Theoritical Framework}

\begin{figure}[!t]
\centering\centering\includegraphics[totalheight=0.8\textwidth]{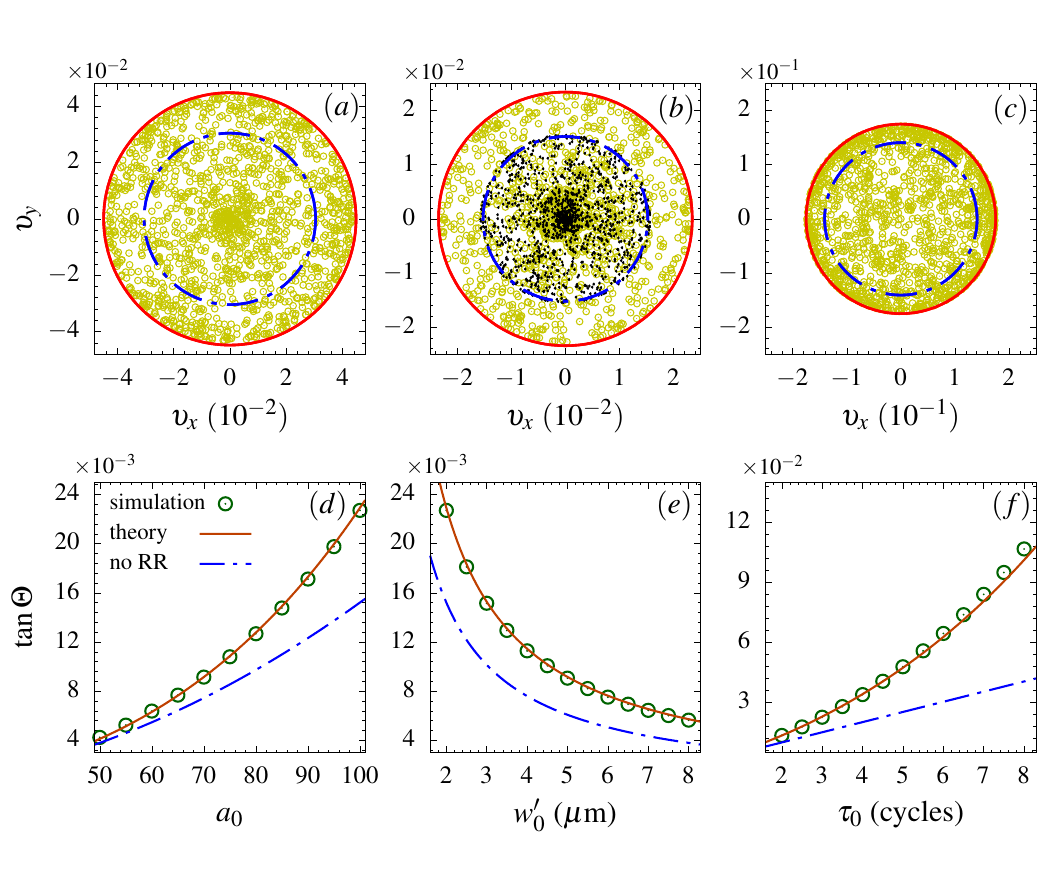}
\caption{The simulated final transverse velocity of outgoing electrons are compared with the predictions of the equation  \ref{thnorr} for various interaction parameters [$a_0$,$\tau_0$ (cycles),$w_0'$($\mu$m),$\gamma_i$]: (a) [100,4,3,200], (b) [100,3,2,300], and (c) [100,5,4,90]. The outer (inner) red (blue) circle represents the theoretically predicted scattering angle when the RR is included (excluded). The black dots in (b) are for the case when the RR is forcefully disabled in the simulation. Next, we present and compare the scaling of the: (d) $a_0$ for a fixed $\tau_0 = 3$ cycles and $w_0' = 2 \mu$m ; (e) $w_0'$ for fixed $a_0 = 100$ and $\tau_0 = 3$ cycles ; and (f) $\tau_0$ for fixed $a_0 = 100$ and $w_0' = 2 \mu$m (refer text for $\gamma_f$ used in these scaling laws). We have used $\gamma_i = 300$ for the results presented in (d), (e) and (f). The points/small circles in all the scaling curves denote the results from the simulation and solid line represent the estimation from the equation  \ref{thnorr}, and dashed blue line is case when RR is disabled in equation  \ref{thnorr} i.e. $\gamma_f = \gamma_i$ is considered.  }
\label{fig2}
\end{figure}

We model the laser field as,
\be  \mb{A} = \frac{a_0}{\sqrt{1 + z^2/z_r^2}}  \exp\left[\frac{-r^2}{w_0^2(1+z^2/z_r^2)}\right] 
g(\eta) \cos(\eta)\ \mb{e_x},  \label{TheovecPot} \ee 
which closely resembles the paraxial beam model used for the numerical results as discussed in \cite{PhysRevSTAB.5.101301}. For now we do not consider RR, such that the relativistic equations of motion of an electron under the influence of these laser fields is given by (using naturalized units $e = m_e = c = 1$),
$d\mb{p}/dt = -\mb{E} - \mb{\upsilon}\times\mb{B}$. The force equation  along the polarization direction can be written as:
\be 
\DDT{\tilde{p}_x} = - \upsilon_x \Pdd{A_x}{x} = \frac{p_x}{\gamma} \frac{2 x}{w_0^2(1 + z^2/z_r^2)} A_x, \label{pxtilde}
\ee
where we noted $A_y = A_z = 0$, $\tilde{p}_x \equiv p_x - A_x$ and $\upsilon_x = p_x/\gamma$ is the electron's velocity along the polarization direction. There will be no external force on the electron once the laser is over or it leaves the focus. Hence in a plane wave laser field it holds $p_x \sim A_x$. In a finite, non-plane-wave field, as considered in equation   \ref{TheovecPot}, however, $\tilde{p}_x$ arises as a correction term due to the gradient of the laser's radial intensity profile. The displacement along the polarization direction can easily be estimated to be $x \sim x_0 + p_x \eta/2\gamma$, with $x_0$ being the initial position of the electron in the bunch.  Using these simple approximations equation   \ref{pxtilde} simplifies to,
\be \DDTa{\tilde{p}_x} \sim \frac{2}{\gamma \DDT{\eta}} \frac{A_x^2}{w_0^2(1 + z^2/z_r^2)} 
\Big[x_0 + \frac{A_x \eta}{2\gamma}\Big] \label{ptildeeta}.\ee
Similarly, the longitudinal component of the force can be written as: $dp_z/dt = - \upsilon_x \pddz{A_x}$, however the rate change of energy is given by $d\gamma/dt = -\upsilon_x \pddt{A_x}$.   It can be shown that in a plane wave $\zeta_z \equiv \gamma - p_z$ is a constant of motion, whose value can be obtained by imposing the condition that initially $p^i_{z} = - \gamma_i \upsilon^i_{z} \sim - \gamma_i$, where $\gamma_i$ is the \textit{initial} energy of the electron bunch and minus indicates the counter-propagation of the electron bunch. Next, using the definition of $\eta = t - z + \phi_0$, one can write $d\eta/dt = 1 - p_z/\gamma$, which implies  $ \gamma d\eta/dt  = \langle\zeta_z\rangle \sim \gamma_i + \gamma_f$. There will be no prominent transverse deflection of the electron until it is near the Rayleigh range of the focused laser. So effectively the $x_0$ can be considered as the initial position of the particle at $z = z_r$. With these approximations, equation   \ref{ptildeeta} is simplified to be
\be \DDTa{\tilde{p}_x} \sim \frac{2}{(\gamma_i + \gamma_f)} \frac{A_x^2}{2 w_0^2} 
\Big[x_0 + \frac{A_x \eta}{2\gamma}\Big] \label{ptildeeta2}\ee   
Terms with $\eta A_x^3$ vanish upon integration whence equation   \ref{ptildeeta2} can be easily integrated over $\eta$ to obtain $\tilde{p}_x$
\be \tilde{p}_x \sim \frac{a_0^2}{(\gamma_i+\gamma_f)}\frac{x_0}{2 w_0^2}\exp\Big[\frac{-r_0^2}{w_0^2}\Big]
\int\limits_{-\infty}^{\infty} \exp\Big[\frac{-8 \ln(2) \eta^2}{\tau_0^2}\Big] \cos^2(\eta) d\eta \label{integration} \ee
\be \tilde{p}_x \sim \sqrt{\frac{\pi}{32\ln(2)}} \frac{a_0^2\tau_0}{(\gamma_i + \gamma_f)}
\frac{x_0}{2 w_0^2}\exp\Big[\frac{-r_0^2}{w_0^2}\Big] \label{pxf0}.\ee
There are manifold scattering angles associated with each individual electron in the bunch, but the width of the electron bunch on the detector, viz. the most easily accessible signal, is set by the largest scattering angle. This largest scattering depends on the maximum transverse momentum. From, equation   \ref{pxf0} it can be deduced that the $\tilde{p}_x$ would maximize for an electron with initial position $x_0 = w_0/\sqrt{2}$ by setting $d\tilde{p}_x/dx_0 = 0$, and hence the maximum net momentum along $x$ can be written as $\tilde{p}_{x;max} = \tilde{p}_x|_{x_0=w_0/\sqrt{2}}$ and is given by,
\be\tilde{p}_{x;max} \sim \sqrt{\frac{\pi}{256\ \ln(2)\ \text{e}}}\ \ \frac{a_0^2}{(\gamma_i +\gamma_f)}\frac{\tau_0}{w_0}.\ee 

The maximum scattering angle of the electron bunch (as estimated from the $z$ axis) would be; $\tan \Theta \propto \tilde{p}_{x;max}/p_z$. As discussed earlier, after the scattering $p_z \sim \gamma_f$ and hence the scattering angle is estimated to be
\be \tan\Theta \sim \kappa \sqrt{\frac{\pi}{256\ \ln(2)\ \text{e}}}\ \ \frac{a_0^2}{\gamma_f(\gamma_i+\gamma_f)}\frac{\tau_0}{w_0}.\label{thnorr}\ee 
where, $\kappa$ is a proportionality constant, to be obtained by fitting the equation   \ref{thnorr} with the simulation results. The fitting parameter $\kappa$ accounts for stronger scattering in a realistic laser field due to higher order field corrections, unaccounted for in our simple model along with other approximations.  This value of the $\kappa$  is obtained by  rigorous benchmarking of the analytical results against our realistic simulations with paraxial laser field profiles.

\subsection{Addressing Radiation Reaction in this model}

\begin{figure}[t]
\centering\includegraphics[totalheight=0.7\columnwidth]{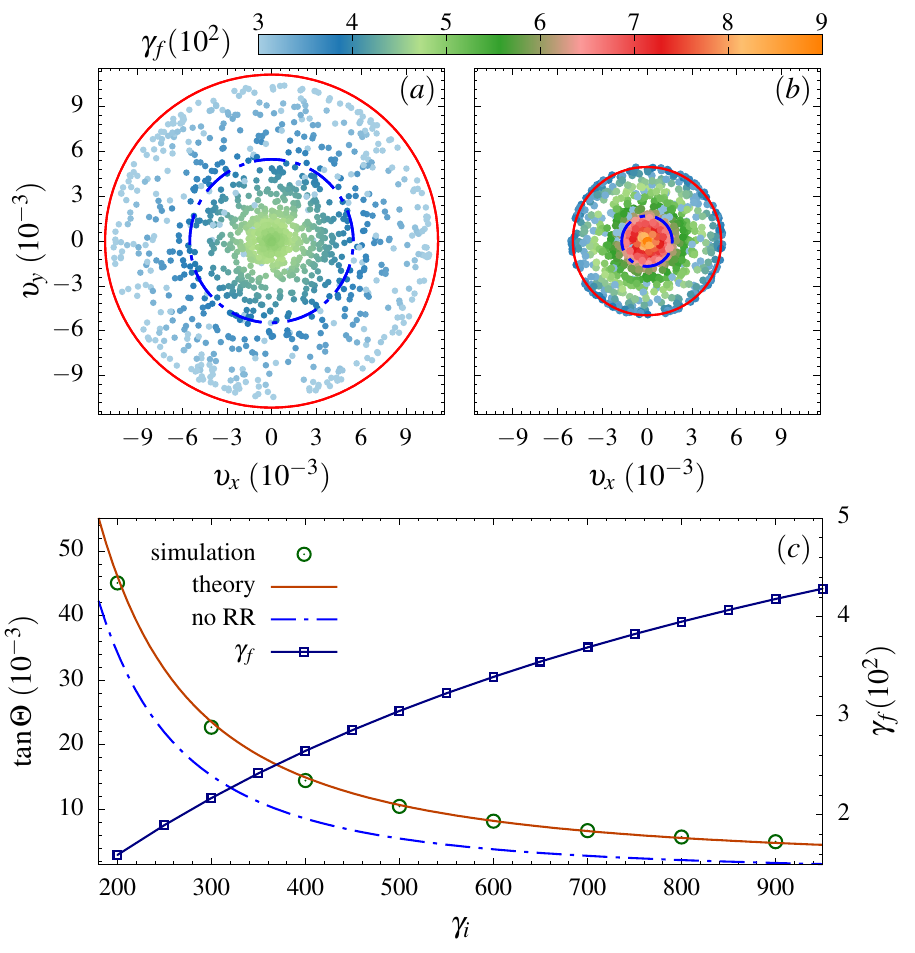}
\caption{The numerically calculated final transverse velocity of electrons for $\tau_0 = 3$ cycles, $a_0 = 100$, and $w_0' = 2 \mu$m are calculated by using $\gamma_i = 500$ (a) and $\gamma_i = 900$ (b). The color coded points in (a) and (b) represent each test particle with their respective $\gamma_f$, and the outer sold red (inner dashed blue) circle is theoretically estimated scattering angle with RR (without RR). The scaling of the scattering angle with the incoming electron bunch energies is presented in (c). Open circles (solid orange line) represent the results from the full test particle simulations (theoretical calculation). The outgoing energy of the single electron ($\gamma_f$) as if it interacts with the plane wave of amplitude $A^\text{p.w.}$ [refer equation  \ref{Apw}] is shown by the line with points on right $y$ axis. These estimates of $\gamma_f$ are actually used in equation  \ref{thnorr} for the scaling (solid orange line) in this figure. The dashed blue line is case when RR is ignored.}
\label{fig3}
\end{figure}

As we have seen, for a focused laser pulse we only require the final energy $\gamma_f$ of the most strongly scattered electrons to estimate the maximal scattering angle of the overall electron bunch. Generally, $\gamma_f \sim \gamma_i$ when there is no loss of energy through RR. However, For ultra-intense laser beams RR implies radiative energy loss of the electrons. Consequently, when RR can not be ignored, the energy of the outgoing electron bunch is always smaller than the energy of the incident electron bunch. In the context of ponderomotive scattering of a relativistic electron bunch, the detailed account of the interaction dynamics is not crucial, but the knowledge of final energy of the electron ($\gamma_f$) after the exit from the laser focus suffices to estimate the extent of the scattering and radiative energy loss. The dominance of RR in the interaction dynamics is conventionally measured in terms of the parameter $R(a_0,\gamma_i) \sim 4 r_{eo} a_0^2 \gamma_i/3$, where $r_{eo} = k r_e \sim 2.21\times 10^{-8}$ (for $\lambda = 800$ nm) is a dimensionless rescaling of the classical electron radius $r_e$. Physically, the parameter $R(a_0,\gamma_i)$ corresponds to the average energy loss due to radiation per laser cycle in units of the electron rest mass energy.   Generally, the quantum effects in RR are prevalent when the quantum efficiency parameter $\chi \sim \gamma_i E_{Laser}/E_S > 1$, where $E_{Laser}$ is the peak laser field and $E_S = m_e^2 c^3/|e|\hbar \simeq 1.32\times 10^{18}$ V/m is the Schwinger critical field. In terms of dimensionless electric field $a_0 = e E/m_e\omega c$, the critical field can be expressed as $a_{S} \sim m_e c \lambda/h \sim 329718$ for $\lambda = 800$ nm laser. For the extreme case of $a_0 = 100$,  $\gamma_i = 900$ used in this work the parameter $\chi \sim 0.27 < 1$ and hence quantum RR would not be crucial for the results presented in our manuscript.

We wish to stress, on the other hand, that the full electron dynamics in a focused laser field with the inclusion of the RR was found too involved for an analytical solution. However, the dynamics of an electron in a plane wave with RR has been extensively studied in the past. Consequently, an accurate estimate of the most strongly scattered electrons' final energy can be found by approximating this final energy as that of an electron scattered by a plane wave with an intensity equal to the intensity an electron would experience inside the focused laser pulse at $(r,z)=(w_0/\sqrt{2},z_r)$. This peak intensity can be read off from equation  \ref{TheovecPot} as  
\be  A^\text{p.w.} = \left|\mb{A}\left(\frac{w_0}{\sqrt{2}},z_r\right) \right| = \frac{a_0}{\sqrt{2}} \exp[-0.25],
\label{Apw} 
\ee 
where we assumed the field to be at the peak of its temporal field oscillations $g(\eta)\cos(\eta)=1$.

The final energy $\gamma_f$ is then estimated by \textit{numerically} solving the LL model dynamics of a single electron inside a plane wave of intensity $A^\text{p.w.}$. The energy loss due to the RR is typically linear with the laser amplitude $a_0$, and the RR effect can be quantified using the dimensionless parameter $R(a_0,\gamma_i)$. In view of this in we fitted the outgoing energy of the electron through the function $\gamma_f(a_0,\tau_0,\gamma_i) = C \sqrt{R(a_0,\gamma_i) \tau_0} + D$ [here $C$ and $D$ are the fitting parameters].  It should be understood that the fitting parameter $C$ and $D$ are not the part of the our analytical framework which predicted the scattering angle as in equation  \ref{thnorr}. However, the final exit energy of the particle is just an input to equation  \ref{thnorr}. There is no straightforward way to calculate $\gamma_f$ (apart from experimentally measuring the same) and hence we relied on the plane wave ansatz and calculated through numerical simulations by fitting the function $\gamma_f(a_0,\tau_0,\gamma_i) = C \sqrt{R(a_0,\gamma_i) \tau_0} + D$. We note, that in a real focused laser field, however, the electron would scatter away, thus experiencing lower RR and hence the final exit energy would be slightly higher than estimated via this simple plane wave model. We will see, however, that this does not affect the predictive power of our model.

\section{Thomson scattered radiation in laser-cluster interaction}

This section is based on our previous work on classical Higher-Harmonic generation from the laser-cluster interaction \cite{PhysRevAccelBeams.22.084401}.  In this case we have used the same laser pulse as used while studying the temporal Gaussian envelope previously. 

 \begin{figure}[t]
\centering\includegraphics[totalheight=3.5in]{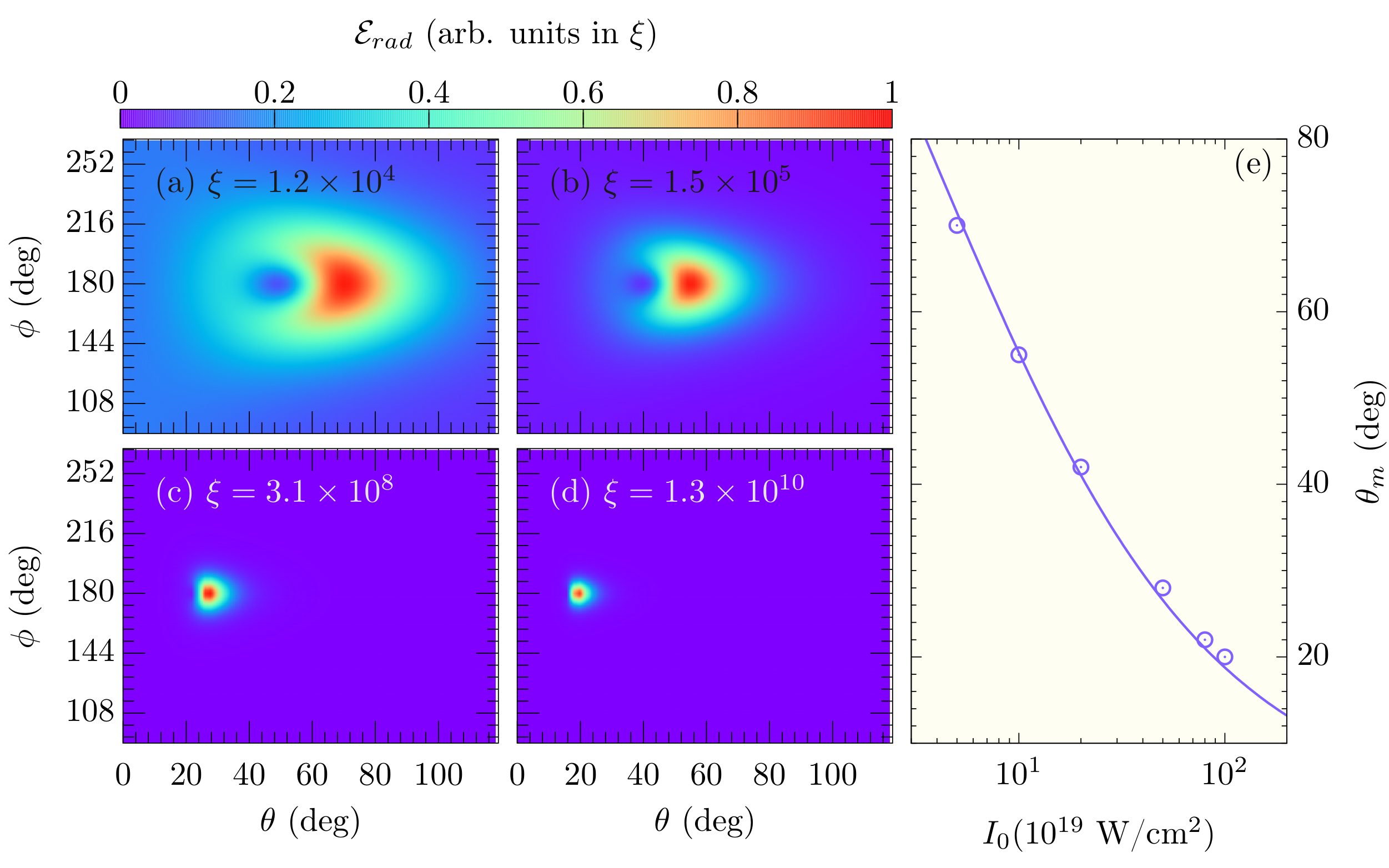}
\caption{Normalized angular distribution (scale factor $\xi$) of the total energy radiated per unit solid angle ($\mathcal{E}_{rad}$) is presented for  (a) $I = 5\times 10^{19}$ W/cm$^2$,  (b) $10^{20}$ W/cm$^2$, (c), $5\times 10^{20}$ W/cm$^2$ and (d) $10^{21}$ W/cm$^2$.  The angle at which maximum energy is detected ($\theta_m$) vs. laser intensity is presented as open circles in (e), and solid line represents the theoretical estimate as given by equation  \ref{angp}.  }
\label{angdist}
\end{figure}

 The trajectory of an electron under the influence of a very intense laser field can be obtained by solving the relativistic equations of motion :
\be \frac{d\mb{P}}{dt} = q [\mb{E} + \mb{v}\times\mb{B}]\ee
where, $\mb{P} = \gamma m_e \mb{v}$ is the relativistic momentum, $\mb{E}$ and $\mb{B}$ are the electromagnetic fields associated with the laser pulse, $q$ and $m_e$ are the charge and mass of the electron. If the laser is polarized along $x$ direction ($\mb{E} = E_x \hat{x}; \mb{B} = B_y \hat{y}$ and $B_y = E_x/c$),  then the dynamics along $x$ and $z$ directions is governed by:
\be\label{vx} \frac{d (\gamma \upsilon_x)}{dt} = \frac{q}{m_e} (E_x - \upsilon_z B_y)  = \frac{q E_x}{m_e } \Big(1 - \frac{\upsilon_z}{c}\Big)\ee
\be\label{vz}\frac{d (\gamma \upsilon_z)}{dt} = \frac{q}{m_e} \upsilon_x B_y  = \frac{q E_x}{m_e }  \frac{\upsilon_x}{c}\ee
The rate change of the energy is given by:
\be\label{gama} \frac{d\gamma}{dt} = \frac{q}{m_e c^2} (\mb{E}\cdot\mb{v}) = \frac{q}{m_e c^2} E_x\upsilon_x\ee
From, equation \ref{vz} and \ref{gama} : 
\be\label{vz0} \frac{d (\gamma \upsilon_z)}{dt} = c \frac{d\gamma}{dt} \implies \frac{\upsilon_z}{c} = 1 - \frac{1}{\gamma}\ee
and from equation \ref{vx}, \ref{gama} and \ref{vz0} we have:
\be \frac{d(\gamma \upsilon_x)}{dt}  = \frac{q E_x}{m_e } \frac{1}{\gamma} = \frac{c^2}{\gamma\upsilon_x} \frac{d\gamma}{dt}\ee
\be (\gamma\upsilon_x)\frac{d(\gamma \upsilon_x)}{dt}  = c^2 \frac{d\gamma}{dt} \implies 
\frac{d}{dt}(\gamma^2\upsilon_x^2 - 2\gamma c^2) = 0 \ee
 by using the initial conditions i.e. $\gamma = 1$ for $\upsilon_x = 0$ we can easily obtain the expression for the $\upsilon_x$ as:
\be\label{vx0} \frac{\upsilon_x}{c} = \frac{\sqrt{2(\gamma-1)}}{\gamma}.\ee
 Furthermore, the relativistic factor $\gamma$ can also be represented as a function of the laser field amplitude as: 
 \be\label{px} \frac{dP_x}{dt} = q E_x - q \upsilon_z B_y = - q\frac{dA_x}{dt} \implies P_x = \gamma m_e \upsilon_x = -q A_x\ee
where, $\mb{A}$ is the vector potential associated with the electromagnetic fields of the laser ($\mb{E} = -\partial \mb{A}/\partial t; \mb{B} = \nabla \times \mb{A}$). Using equation \ref{vx0} and \ref{px} we can obtain the expression for relativistic factor as:
\be\label{g0} \gamma = 1 + \frac{q^2 A_x^2}{2 m_e^2 c^2} \equiv 1 + \frac{a_0^2}{2}\ee 
here, $a_0$ is the dimensionless laser amplitude defined previously. From equation \ref{vx0}, \ref{vz0} and \ref{g0}, the velocity components then can be written as:
\be\label{vel} \frac{\upsilon_x}{c} = \frac{2a_0}{2+a_0^2}\quad;\quad \frac{\upsilon_z}{c} =   \frac{a_0^2}{2+a_0^2}.\ee
The strong ponderomotive push of laser along the propagation direction can easily be estimated by equation \ref{vel} in 
terms of the angle ($\theta_{ponder}$) it makes with the $z$ axis  \cite{Lee-2003}:
\be\label{pond} \theta_{ponder} = \tan^{-1}\Big(\frac{\upsilon_x}{\upsilon_z}\Big) = \tan^{-1}\Big(\frac{2}{a_0}\Big) \ee
The angle at which maximum radiation ($\theta_{lw}$) is emitted when velocity and acceleration are along the same direction 
can be estimated by the potentials:
\be\label{lwp} \theta_{lw} = \cos^{-1} \Big( \frac{\sqrt{1+15\beta^2} - 1}{3\beta} \Big)   \ee
here, 
$$\beta \equiv  \sqrt{1 - \frac{1}{\gamma^2}} = \sqrt{1 - \frac{1}{(1+ 0.5 a_0^2)^2}} $$ 
is the velocity of charged particle and $\theta_{lw}$ is the angle it makes with the velocity 
vector. Now, from equation \ref{pond} we can estimate the direction of electron motion with respect to the propagation axis, and 
equation \ref{lwp} would predict the angle (with respect to the velocity vector) at which maximum radiation is emitted and as a result, 
the estimated angle at which maximum radiation is emitted is given by: 
\be \theta_m = \theta_{ponder} + \theta_{lw}\ee
\be\label{angp} \theta_m = \tan^{-1}\Big(\frac{2}{a_0}\Big) +  \cos^{-1} \Big( \frac{\sqrt{1+15\beta^2} - 1}{3\beta} \Big).   \ee

\newpage
\section{Discussion on LEADS - C++ Code} 
We have created a C++ code named ``\textbf{LEADS}" (Laser Electron interAction Dynamics Simulator) and here the main Physics engine of the code is discussed. 

\subsection{Main file: ``LEADS.cpp"}
\begin{lstlisting}
  /*
 LEADS : Laser Electron interAction Dynamics Simulator
 Amol R Holkundkar,
 Department of Physics,
 Birla Institute of Technology and Science - Pilani, Pilani. 
 India.  
 */

// To compile use the following command 
// g++ LEADS.cpp -lgsl -lgslcblas -lm

#include "LEADS.h" 

// Boris Particle Push Algorithm
// Filippychev D S 2001 Comput. Math. Model. 12 193
// https://link.springer.com/article/10.1023/A:1012589205469

/*********************************************/
int    RR_Enable = 1; // enable Radiation Reaction  
double gama0     = 100;// initial gama of electron

double I0        = 1.0e18; // W/cm^2
double lambda    = 0.8e-6;

//double a0        = sqrt(I0 * SQ(lambda/1e-6) / 1.36816e+18) ; 
double a0         = 40; 
 
double WAVE_FLAG = 1; // 1 - Gaussian ; 0 - Infinite wave
double pola      = 0; // 0 - Linear, 1 - Circular
double delta     = (pola > 0.5 ? 1.0/sqrt(2) : 1) ; 
double d         = 3.0 * 2*M_PI;// Dimensionless pulse length
double TauDL     = 3*d;    // Simulation time in dimensionless units
double PHASE     = -TauDL;

double omega0    = 2*M_PI*c/lambda;
double k         = 2*M_PI/lambda;

double dTau      = 1e-4; //in dimensionless units = Omega*time
 
int dataSTORE    = 10 ;  //data store interval
double RR_CONST_LL =  2 * k * re/3.0 ;
 /*********************************************/

void Initialize_Particle(Particles &par) 
{
  par.gama = gama0;
  par.v0 = sqrt(1 - 1/SQ(par.gama)) ;
  par.m = 1;
  par.q = -1;
  par.x =  0.0;
  par.y =  0.0;
  par.z =  0.0;
  par.vx = 0;par.vy = 0.0;par.vz = -par.v0; // vz in units of c 
  par.px = par.gama*par.vx; // px,py,pz in units of m*c
  par.py = par.gama*par.vy;
  par.pz = par.gama*par.vz; 
  par.ax = 0;par.ay = 0;par.az = 0; 
} 
  
VD EM_Fields :: E_Field(double t,double r)
{
 
 double Ex,Ey,Ez,E0;
 double alpha,eta;
 
 alpha =  4*log(2.0)/SQ(d) ;
 eta = t - r + PHASE;
 
 E0 = a0;
 
 if(WAVE_FLAG > 0)
 {  
  Ex = E0 * delta * exp(-alpha*SQ(eta)) 
       * ( sin(eta) + 2 * alpha * eta * cos(eta) ) ; 
  Ey = E0 * sqrt(1 - SQ(delta)) * exp(-alpha*SQ(eta)) 
       * ( 2*alpha*eta*sin(eta) - cos(eta) );
 }
 else
 {
  Ex = E0 * delta * sin(eta) ; 
  Ey = -E0 * sqrt(1 - SQ(delta)) *  cos(eta) ;
 } 
  
 VD field;
 field.assign(4,0.0);
 
 field[1] = Ex ;
 field[2] = Ey ;
 field[3] = Ez ;
 
 return field;
}


VD EM_Fields :: B_Field(double t,double r)
{
 
 double Bx,By,Bz,B0;
 double alpha,eta;
 
 alpha =  4*log(2) /SQ(d) ;
 eta = t - r + PHASE ;
  
 B0 = a0;
 
 if(WAVE_FLAG > 0)
 { 
  Bx = -B0*sqrt(1 - SQ(delta)) * exp(-alpha*SQ(eta)) 
        * ( 2*alpha*eta*sin(eta) - cos(eta) ); 
  By = B0 * delta * exp(-alpha*SQ(eta)) 
        * ( sin(eta) + 2 * alpha * eta * cos(eta) ) ; 
 }
 else
 { 
  Bx = B0*sqrt(1 - SQ(delta))  * cos(eta) ; 
  By = B0 * delta *  sin(eta)  ; 
 }
   
 VD field;
 field.assign(4,0.0);
 
 field[1] = Bx ;
 field[2] = By ; 
 field[3] = Bz ;
 
 return field;
}

int main()
{
  ofstream outf("data.dat",ios::out); // data file
  outf << scientific << uppercase << setprecision(6);
 
  cout << endl << "delta = " << delta;  
  cout << endl << "a0    = " << a0;
  cout << endl << "gama0 = " << gama0;
  cout << endl << "TauDL = " << TauDL;
  cout << endl << "dTau  = " << dTau;
  
  Initialize_Particle(ELEC);
   
  VD e_field;
  int counter = 0;
 
  // Main time loop 
  for(double tr = 0;tr< TauDL;tr+=dTau)
  {
   
   CalForceMovePart(tr); // calculate force and move particles
 
   if(counter/dataSTORE == 1)
   {
    e_field = em.E_Field(tr,ELEC.z) ;
    outf << endl << tr << setw(15) << tr - ELEC.z << setw(15) 
    << e_field[1] << setw(15) << e_field[2] << setw(15) 
    << e_field[3] << setw(15) 
    << ELEC.x << setw(15) << ELEC.y << setw(15) << ELEC.z 
    << setw(15)  
    << ELEC.vx << setw(15) << ELEC.vy << setw(15) << ELEC.vz 
    << setw(15) << ELEC.gama; 
 
    counter = 0;
   }
   counter++;
  }
   
 outf.close();
 cout << endl;
 return 0;
}
  
void CalForceMovePart(double Time)
{
   
 double func,gama,gamaOld;
 double uxb1,uxb2,uxb3,udb;
 double del,del_s;
 double Uxm,Uym,Uzm;
 double Uxp,Uyp,Uzp;
 double B2,del2;
 double term1,term2;
 double r;
 VD vxb(4),Exv(4);
 VD vv(4);
 VD bb;
 VD ee;
 VD wB;
 vxb.assign(4,0.0);vv.assign(4,0.0);
 Exv.assign(4,0.0);
 bb.assign(4,0.0);
 Vector P,UP,UM,UxB,E,B,ACC;
 Vector Pold,Pave,Vold ;
  
 bb = em.B_Field(Time,ELEC.z); 
 ee = em.E_Field(Time,ELEC.z); 
 E = assign(ee[1],ee[2],ee[3]);
 B = assign(bb[1],bb[2],bb[3]);
 
 Vold = assign(ELEC.vx,ELEC.vy,ELEC.vz);
 P = assign(ELEC.px,ELEC.py,ELEC.pz);
 ACC = assign(ELEC.ax,ELEC.ay,ELEC.az); // acceleration
 Pold = P ;
 
 gama = sqrt(1 + SQ(P.x) + SQ(P.y) + SQ(P.z)) ;
 ELEC.gama = gama;
 gamaOld = gama;
 
 del_s = (ELEC.q/ELEC.m) *0.5 * dTau;
 del = del_s/gama;
 
 B2 = SQ(vecMod(B)) ;
 del2 = del*del;
 
 term1 = (1 - del2*B2) / (1 + del2*B2);
 term2 =  2*del / (1 + del2*B2);
 
 UM = add(P,scalar(E,del_s)) ;
 UxB = cross(UM,B);
 udb = dot(UM,B);
 UP = add(add(scalar(UM,term1),scalar(UxB,term2)),scalar(B,del*term2*udb) ) ;
 P = add(UP,scalar(E,del_s));
 
 if (RR_Enable == 1) // Landau-Lifshitz Model
 {
 
 //Radiation Reaction Part
 //New Journal of Physics 12 (2010) 123005, 
 //"Radiation reaction effects on radiation pressure acceleration", 
 //M Tamburini et al
 
 Vector V,FL,FR,FR1,FR2,PR;
 double VdotE,FL2;
 V = scalar(Pold,1/gamaOld);
 FL = scalar(add(E,cross(V,B)), ELEC.q ) ; 
 FL2 = dot(FL,FL);
 VdotE = dot(V,E);
 FR1 = add(cross(B,FL),scalar(E,VdotE)); // B x FL = - FLx B 
 FR2 = scalar(V,- SQ(gamaOld)*(FL2 - SQ(VdotE))) ;
 FR = scalar(add(FR1,FR2),RR_CONST_LL); 
 PR = add(scalar(FR,dTau),Pold);
 
 ELEC.px = P.x + ( PR.x - Pold.x );
 ELEC.py = P.y + ( PR.y - Pold.y );
 ELEC.pz = P.z + ( PR.z - Pold.z );
 
 ELEC.rr0 = vecMod(FR);
 ELEC.rr1 = 0;
 ELEC.FL = FL;
 //------------------------------------------------------
}
else
{
 ELEC.px = P.x;
 ELEC.py = P.y ;
 ELEC.pz = P.z;
}
   
 gama = sqrt(1 + SQ(ELEC.px) + SQ(ELEC.py) + SQ(ELEC.pz)) ;

 ELEC.vx = ELEC.px / gama;
 ELEC.vy = ELEC.py / gama;
 ELEC.vz = ELEC.pz / gama;
  
 ELEC.x = ELEC.x + ELEC.vx * dTau ;
 ELEC.y = ELEC.y + ELEC.vy * dTau ;
 ELEC.z = ELEC.z + ELEC.vz * dTau ; 
 
 ELEC.ax = (ELEC.vx - ELEC.vx0)/dTau;
 ELEC.ay = (ELEC.vy - ELEC.vy0)/dTau;
 ELEC.az = (ELEC.vz - ELEC.vz0)/dTau;
 
 ELEC.vx0 = ELEC.vx;
 ELEC.vy0 = ELEC.vy;
 ELEC.vz0 = ELEC.vz;
 
}
 

 \end{lstlisting}

\subsection{Header file: ``LEADS.h"}

\begin{lstlisting}
#include <iostream>
#include <cmath>
#include <fstream>
#include <vector>
#include <stdio.h>
#include <cstring>
#include <sstream>
#include <cstdlib>
#include <iomanip>
#include <gsl/gsl_spline.h>
 
#define c 2.99792458E+8 // Speed of light in meters/second
#define e 1.602176E-19	// Charge of electron in Coulombs
#define me 9.109382E-31 // Mass of electron in kg
#define re 2.8179402894E-15
#define ns 1.0E-9	// Nanosecond
#define nm 1.0E-9	// Nanometer
#define Ang 1.0E-10     // Angstrom
 
double SQ(double x) {return x*x;}
double CUBE(double x) {return x*x*x;}

using namespace std;
typedef vector <double>  VD;

class EM_Fields
{
public:
        VD E_Field(double,double);
        VD B_Field(double,double); 
        VD A_Field(double,double); 
};
EM_Fields em;

void CalForceMovePart(double); // Calculate force and move particle
  
/*********************************************/
// For vector Algebra
struct Vector
{
 double x,y,z;
};

Vector assign(double x,double y,double z)
{
 Vector dum0;
 dum0.x = x;
 dum0.y = y;
 dum0.z = z;
 return dum0;
} 

//Defining particles 
struct Particles
{
 double m,q,gama,v0;
 double x,y,z;
 double px,py,pz;
 double vx,vy,vz;
 double vx0,vy0,vz0;
 double ax,ay,az;
 double KineEner;
 int escapeID;
 double rr0,rr1; // Radiation reaction terms
 Vector FL;
};
Particles ELEC;

double dot(Vector A,Vector B)
{
 return A.x*B.x + A.y*B.y + A.z*B.z;
}
 
double vecMod(Vector A)
{
 return sqrt(A.x*A.x + A.y*A.y + A.z*A.z);
}

Vector cross(Vector A,Vector B)
{
 Vector C;
 C.x = A.y * B.z - A.z * B.y;
 C.y = A.z * B.x - A.x * B.z ;
 C.z = A.x * B.y - A.y * B.x;
 return C;
}

Vector scalar(Vector A,double a)
{
 Vector C;
 C.x = A.x * a;
 C.y = A.y * a;
 C.z = A.z * a;
 return C;
}


Vector add(Vector A,Vector B)
{
 Vector C;
 C.x = A.x + B.x;
 C.y = A.y + B.y;
 C.z = A.z + B.z;
 return C;
}

Vector sub(Vector A,Vector B)
{
 Vector C;
 C.x = A.x - B.x;
 C.y = A.y - B.y;
 C.z = A.z - B.z;
 return C;
}
/*********************************************/



\end{lstlisting}

\section*{Acknowledgement}
The author acknowledges the use of ChatGPT by OpenAI for content generation in preparing this article. The tool is only used for the content which is well established academically. This tool is also used to generate {\LaTeX} code from the personal Hand-written notes. The generated content has been carefully curated and is accurate to the best of the author's knowledge.

\bibliographystyle{aipnum4-1}
%
 
\end{document}